\RequirePackage{fix-cm}
\documentclass[twocolumn]{svjour3}          
\smartqed  
\usepackage{graphicx}
\usepackage{colortbl}
\usepackage{mathptmx}      
\usepackage{latexsym}
\usepackage{marvosym}
\usepackage{natbib}
\usepackage{amsmath}
\usepackage{amssymb}
\begin{document}


\title{Aspects of randomness in neural graph structures}

\titlerunning{Randomness in neural graphs}

\author{Michelle Rudolph-Lilith         
        \and
        Lyle E. Muller
}

\institute{M. Rudolph-Lilith (\Letter) \and L.E. Muller 
           \at
              Unit\'e de Neurosciences, Information et Complexit\'e (UNIC) \\
              CNRS, 1 Ave de la Terrasse, 91198 Gif-sur-Yvette, France \\
              Tel.: +33-1-69-82-41-73 \\
              Fax: +33-1-69-82-34-27 \\
              \email{rudolph@unic.cnrs-gif.fr}
}

\date{Received: date / Accepted: date}

\maketitle


\begin{abstract}

In the past two decades, significant advances have been made in understanding the structural and functional properties of biological networks, via graph-theoretic analysis. In general, most graph-theoretic studies are conducted in the presence of serious uncertainties, such as major undersampling of the experimental data. In the specific case of neural systems, however, a few moderately robust experimental reconstructions do exist, and these have long served as fundamental prototypes for studying connectivity patterns in the nervous system. In this paper, we provide a comparative analysis of these ``historical'' graphs, both in (unmodified) directed and (often symmetrized) undirected forms, and focus on simple structural characterizations of their connectivity. We find that in most measures the networks studied are captured by simple random graph models; in a few key measures, however, we observe a marked departure from the random graph prediction. Our results suggest that the mechanism of graph formation in the networks studied is not well-captured by existing abstract graph models, such as the small-world or scale-free graph.



\end{abstract}


\section{Introduction}
\label{S_Introduction}

Since Stanley Milgram's six degrees of separation \citep{Milgram67}, characterization of topological structure has become a major focus of graph-theoretic investigations in complex networks \citep{CostaEA05}. In recent years, studies of this kind have begun to play an important role in a wide variety of disciplines, ranging from communications and power systems engineering, to molecular and population biology \citep{AlbertEA99, AlbertBarabasi02, DorogovtsevMendes02, AlmArkin03, Alon03, Bray03, Newman03, BarabasiOltvai04}. Often, by applying simple graph-theoretical measures, it is possible to find similarities in real-world graphs describing systems in many different domains, and also to separate these graphs into a number of representative classes, by highlighting their differences. From this, several studies have moved forward to connect such shared similarities to abstract, theoretical models of graph generation, which in turn can then be used to further investigate real-world graphs beyond the limitations imposed by the technologies currently available.

Two of the most successful of these models are the small-world and scale-free graphs \citep[for a general review see \citealp{BoccalettiEA06, Newman10}]{WattsStrogatz98, AlbertEA99}. In particular, scale-free graphs are generally viewed as a crucial prerequisite for complex dynamical behaviors, and have been identified as a unifying feature of many real-world graphs \citep{BarabasiBonabeau03, AmaralOttino04}. In recent years, however, several studies have challenged the empirical support for scale-free properties in many real-world graphs, and their mechanistic backing \citep{ClausetEA09, LimaMendezVanHelden09, StumpfPorter12}. There is a growing consensus that the evidence for scale-free properties needs to be carefully re-considered. The insights gained from this may in turn lead to a deeper understanding of the underlying mechanisms producing the large-scale structure seen in real-world systems.  

Among the most challenging real-world systems for graph-theoretic characterization are the strongly interconnected networks of the nervous system. Here, the analysis of the structural makeup of these graphs has shown no consistent evidence for scale-free properties. One of the reasons for this lack could be the severe undersampling due to technical limitations in the experimental reconstructions. A few moderately robust experimental reconstructions do exist, however, both of neural connectivity graphs \citep{VarshneyEA11} and areal connectivity maps \citep{ModhaaSingh10}. These data suggest that scale-free organization is rather unlikely, as the number of connections per graph node generally does not span multiple scales. For instance, it is known in the mammalian cortex that the typical number of synaptic connections for a single neuron varies over one or, at most, two orders of magnitude \citep{BraitenbergShuz98}, ruling out the possibility of power-law organization over multiple scales in this structural network. While this argumentation does not necessarily apply to functional brain networks, and there has been evidence recently presented for their scale-free organization (e.g., see \citealp{EguiluzEA05}), other studies have report conflicting observations \citep{LimaMendezVanHelden09, StumpfPorter12}, and this remains an open question (for review, see \citealp{BullmoreSporns09}). 

In this study, we provide a comparative analysis of several ``historical'' reconstructions of structural neural graphs, including areal connectivity maps of the cat and macaque monkey cortex, as well as the neural connectivity graph of the nematode {\it Caenorhabditis elegans}. The general subject of interest here is the assessment of the structural connectivity pattern in these graphs. To this end, we make use of a set of simple measures, which characterize various aspects of the connectivity pattern. Specifically, we consider the node degree distributions, the structural equivalence of graph nodes, as well as a nearest neighbor degree and assortativity. Throughout this work, all measures are defined in their most general fashion, for directed graphs, but yield their forms known from the literature when applied to undirected graphs. By applying the same measures to both the original directed and symmetrized undirected versions of each considered graph, we demonstrate that the process of symmetrization not only places limits on the characterization of graphs, but also introduces a systematic bias in measurement.

We find that the investigated networks share a strong component of randomness in their structural makeup, suggesting a mechanism of their formation which is much less constrained than that required for scale-free graphs. However, the observed graph structures differ from that of the Erd\H{o}s-R\'enyi random graphs most widely used in computational and theoretical studies of neural networks, by their specific node degree distribution and strong correlations of in-coming and out-going connections for individual nodes.


\section{Methods}


\subsection{Graph theory preliminaries}

A graph or network is comprised of a set of nodes which are linked by a set of edges. Two general types of graphs can be distinguished: undirected graphs, for which all edges act as bidirectional links between two nodes, and directed graphs (digraphs), in which case each edge is endowed with a direction pointing from a source node to a target node. In both cases, the spatial position of nodes can be considered (spatial graphs), and edges can exhibit properties such as a delay (delayed graphs) or weight (weighted graphs). Relational graphs are those excluding these additional properties, taking only the relations, or adjacencies, between nodes into account. 

In this work, only relational graphs will be considered. In this case, the relationship between nodes can be mathematically formulated using an adjacency matrix $a_{ij}$, $i,j \in [1, N_N]$, where $N_N$ denotes the number of nodes in the given graph. If node $i$ makes a connection to node $j$, then $a_{ij} = 1$, otherwise $a_{ij} = 0$. Undirected graphs are a special case of digraph with symmetric adjacency matrix, i.e.~$a_{ij} = a_{ji}$.

Special attention is required for the diagonal elements $a_{ii}$ of the adjacency matrix, which describe self-loops. In the case of digraphs, a self-looped node acts both as target and source, so that non-zero diagonal elements contribute always two edges to a graph. For undirected graphs, this also leads to the contribution of two edges per self-looped node. This definition, however, deviates from that commonly used, because historically, undirected graphs were considered first, and most notions in modern graph theory will use only one edge per undirected link. In this paper we will follow the latter notion with the exception of self-loops, which contribute two edges, and present all measures utilized in both their undirected and directed form.

For a general introduction to graph theory, its measures and applications, we refer to \citep{Diestel00, BoccalettiEA06, Newman10}.


\subsection{Analysis methods}

In this study, we used graph data which are publicly available. No modifications of the original data were performed. However, some of the graphs experienced a certain level of modifications since their first investigation. Therefore, numerical results reported in this study might deviate slightly from results reported in earlier studies.

It is a common practice in applying graph-theoretic measures to empirical data to symmetrize the adjacency matrix prior to analysis. Throughout this paper, both the original, directed graphs and their undirected, symmetrized versions are considered. Undirected graphs were symmetrized by setting each pair $(i, j)$ with $a_{ij} = 1$, $a_{ji}$ to 1. The reduction to the giant component and subsequent analysis was performed on the symmetrized versions of the original graphs.

In some cases, we constructed corresponding Erd\H{o}s-R\'enyi graph models for the given graphs with the same number of nodes, edges and node degree distributions (node in/out-degree distributions for digraphs). These exact degree-matched Erd\H{o}s-R\'enyi graph models (EDM) were obtained using the {\it cygraph} implementation of a sophisticated model introduced in \citet{DelHenioEA10} and \citet{KimEA12}. When considering EDM graphs, 1,000 random realizations were used for each parameter set to ensure statistical stability. 

Numerical analysis was performed using the custom software {\it cygraph} and {\it Mathematica}. A {\it cygraph} binary (Mac OSX), all graph data and analysis protocols are available for download\footnote{http://www.cydyns.com; http://www.newscienceportal.com/MLR}.

\begin{table}[b]
\caption{Basic graph-statistical measures were applied to the original directed and symmetrized (undirected) versions of the considered neural graphs. Shown are values for the number of nodes $N_N$, number of self loops $N_L$, total adjacency $A$ (Eq.~\ref{Eq_TotalAdjacency}; $N_E$ given in Eq.~\ref{Eq_NumberOfEdges}), asymmetry index $\mathcal{A}$ (Eq.~\ref{Eq_AsymmetryIndex}; $\mathcal{A} = 0$ for undirected graphs) and connectedness $Co$ (Eq.~\ref{Eq_Connectedness}).}
\label{Tab_BasicProperties}
\begin{tabular*}{\columnwidth}{l|p{2mm}p{2mm}|lll|p{6mm}l}
\hline
     & & & \multicolumn{3}{c|}{{\bf directed}} & \multicolumn{2}{c}{{\bf undirected}} \\
     & $N_N$ & $N_L$ & $A$ & $\mathcal{A}$ & $Co$ & \multicolumn{1}{c}{$A$} & $Co$ \\
\hline
CC1  &  95 & 0 & 2126 & 0.1829 & 0.2331 &  2340 & 0.2566 \\
CC2  &  52 & 0 &  818 & 0.4117 & 0.2968 &  1030 & 0.3737 \\
CE1  & 306 & 0 & 2345 & 0.9083 & 0.0250 &  4296 & 0.0457 \\
CE2  & 297 & 0 & 2345 & 0.9083 & 0.0265 &  4296 & 0.0486 \\
CE3  & 279 & 3 & 2996 & 0.6917 & 0.0384 &  4580 & 0.0586 \\
MB1  & 383 & 0 & 6602 & 0.7323 & 0.0449 & 10416 & 0.0708 \\
MC1  &  71 & 0 &  746 & 0.2968 & 0.1459 &   876 & 0.1714 \\
MC2  &  94 & 0 & 2390 & 0.4224 & 0.2676 &  3030 & 0.3393 \\
MNC1 &  47 & 0 &  505 & 0.3866 & 0.2238 &   626 & 0.2775 \\
MVC1 &  30 & 0 &  311 & 0.3632 & 0.3344 &   380 & 0.4086 \\
MVC2 &  32 & 0 &  315 & 0.3763 & 0.2983 &   388 & 0.3674 \\
\hline
\end{tabular*}
\end{table}


\subsection{``Historical'' biological neural graphs}

We study structural aspects of a number of publicly available biological neural graphs used in the literature in the past two decades. These include areal connectivity graphs of the cat and macaque monkey cortex, as well as the neuronal connectivity graph of the nematode {\it C. elegans}. Below we briefly describe these graphs, and the notation used throughout this paper. More information on the data sources can be found in the references. 

\subsubsection*{Cat neural graphs} 

The first set contains the areal connectivity graph including all cortical and thalamic areas of the cat brain (CC1), and a graph containing only the 52 cortical areas (CC2). Structural connection data for both CC1 and CC2 were first reported in \citet{ScannellEA99}, and obtained by analyzing a large collection of individual connection tracing studies done in the cortical and thalamic nuclei of the cat cerebral hemisphere. Available connection matrices\footnote{\label{fnote1}https://sites.google.com/site/bctnet/datasets} describe graphs containing 95 nodes and 2126 directed edges (CC1), and 52 nodes and 818 directed edges (CC2). Both graphs were studied in detail in \citet{SpornsZwi04} and \citet{SpornsKotter04}.

\subsubsection*{\textit{C. elegans} neural graphs} 

Three variations of the neuronal connectivity graph of the nematode worm {\it C. elegans} most often used throughout the literature were studied. Data for the first two (CE1 and CE2) are based on experimental data from \citet{WhiteEA86}, and were modified and made public in \citet{WattsStrogatz98}. Available connection matrices\footnote{CE1: http://wiki.gephi.org/index.php/Datasets; CE2: http://www-personal.umich.edu/$\sim$mejn/netdata/ with modifications by M. Newman} describe graphs containing 306 nodes and 2345 directed edges (CE1), and 297 nodes with 2345 directed edges (CE2).

The third dataset (CE3) constitutes the most recent and complete connectivity graph of {\it C. elegans} and was first discussed in \citet{ChenEA06} (for a comprehensive review, see \citealp{VarshneyEA11}). The available connection matrix\footnote{http://wormatlas.org/neuronalwiring.html} describes a graph containing 279 nodes and 2996 directed edges.

All graph data describe the synaptic connections between neurons of the {\it C. elegans} brain, with distinction of directed chemical synapses and undirected electrical junctions. In this paper, we will not consider this distinction, but view both chemical synapses and electrical junctions as part of the same connectivity structure (see \citealp{VarshneyEA11, RudolphLilithEA12} for an analysis of both subgraphs).

\subsubsection*{Macaque monkey neural graphs} 

Various graphs of the macaque brain were considered. The most complete dataset (MB1) describes the macaque brain's long-distance areal connections, and was first described in \citet{ModhaaSingh10}. The obtained connectivity data\footnotemark[2] were assembled from the {\it Collation of Connectivity data on the Macaque brain (CoCoMac)} database. The latter is a growing collection of annotated information about a large number of published tracing studies performed in the macaque  brain \citep{StephanEA01, Kotter04}. The investigated graph contains 383 nodes describing the various brain regions of the macaque monkey, and 6602 directed edges.

A second graph (MC1) describes the areal connectivity pattern of the macaque cortex, based on original data published in \citet{Young93}, and investigated in detail in \citet{SpornsTononi02} and \citet{Sporns02}. The available connection matrix\footnotemark[2] describes a graph containing 71 nodes and 746 directed edges.

A third graph (MC2) contains the the macaque cortical connectivity within one hemisphere, based on data from 
\citet{ChoeEA04}, \citet{Kotter04} and \citet{KaiserHilgetag06}. The available dataset\footnote{http://www.biological-networks.org/?page\_id=25} describes a graph containing 94 nodes and 2390 directed edges.

Finally, we analyzed the visual and sensorimotor area corticocortical connectivity graph of the macaque neocortex (MNC1) and two areal connectivity graphs of the macaque visual cortex (MVC1, MVC2). The MNC1 graph was first studied and made public in \citet{HoneyEA07}. The available dataset\footnotemark[2] describes a graph containing 47 nodes and 505 directed edges. MVC1 and MVC2 are two variants of the visual cortical connectivity originally published in \citet{FellemanVanEssen91}, and investigated in detail in \citet{SpornsEA00} and \citet{SpornsKotter04}. The available datasets\footnotemark[2] describe graphs containing 30 nodes and 311 directed edges (MVC1), and 32 nodes with 315 directed edges (MVC2). 

\begin{table}[b]
\caption{Basic node degree analysis of original directed and symmetrized (undirected) versions of the considered neural graphs. Listed are the minimum and maximum node degree $\delta$ and $\Delta$, respectively ($\delta^{\alpha}$ and $\Delta^{\alpha}$ for digraphs; $\alpha \in \{ in,out \}$), and the average node degree $\langle a_i \rangle$ (Eq.~\ref{Eq_AverageNodeDegree}).}
\label{Tab_NodeDegreeProperties}
\begin{tabular*}{\columnwidth}{l|lllll|lll}
\hline
     & \multicolumn{5}{c|}{{\bf directed}} & \multicolumn{3}{c}{{\bf undirected}} \\
     & $\delta^{in}$ & $\Delta^{in}$ & $\delta^{out}$ & $\Delta^{out}$ & $\langle a_i \rangle$ & $\delta$ & $\Delta$ & $\langle a_i \rangle$ \\
\hline
CC1  & 2 &  55 & 2 &  52 & 22.38 & 2 &  61 & 24.63 \\
CC2  & 7 &  32 & 3 &  34 & 15.73 & 7 &  37 & 19.81 \\
CE1  & 0 & 134 & 0 &  39 &  7.66 & 0 & 134 & 14.04 \\
CE2  & 0 & 134 & 0 &  39 &  7.90 & 1 & 134 & 14.46 \\
CE3  & 0 &  83 & 0 &  57 & 10.73 & 2 &  93 & 16.41 \\
MB1  & 0 & 105 & 0 & 109 & 17.24 & 0 & 149 & 27.20 \\
MC1  & 0 &  26 & 1 &  28 & 10.51 & 1 &  28 & 12.34 \\
MC2  & 0 &  73 & 1 &  54 & 25.43 & 1 &  74 & 32.23 \\
MNC1 & 1 &  23 & 2 &  23 & 10.74 & 3 &  27 & 13.32 \\
MVC1 & 2 &  19 & 4 &  20 & 10.37 & 5 &  22 & 12.67 \\
MVC2 & 0 &  19 & 2 &  20 &  9.84 & 2 &  22 & 12.13 \\
\hline
\end{tabular*}
\end{table}

The basic graph-theoretic properties of these graphs are listed in Tables \ref{Tab_BasicProperties} and \ref{Tab_NodeDegreeProperties}, and further discussed below.


\subsection{Connected components}

A (strongly) connected component is defined as a subgraph consisting of a set of nodes from which all other nodes in the subgraph can be reached, and which can be reached from all other nodes, by following existing edges. Typically, the set of (strongly) connected components of a graph will be dominated by a giant (strongly) connected component of size $S_{gcc}$, defined as the number of nodes in this component \citep{BoccalettiEA06}. We calculated the number of connected components (strongly connected components, in the case of digraphs) $N_{cc}$ and the size of the giant connected component (giant strongly connected component for digraphs) $S_{gcc}$.

Table~\ref{Tab_ConnectedComponents} summarizes the numerical results for the considered biological neural graphs and their symmetrizations, along with the asymmetry index, minimal, maximal and average node degrees, $\delta$, $\Delta$ and $\langle a_i \rangle$, respectively. Naturally, the connectedness of the giant connected component is slightly larger than that of the original graphs, whereas the asymmetry index $\mathcal{A}$ is slightly smaller for graphs whose size of the giant connected component is smaller than the $N_N$ of the original graph. 

Throughout this work, we restrict our analysis to the giant connected (for undirected versions of the considered graphs) and giant strongly connected (for the original digraphs) components. Furthermore, as indicated in Table~\ref{Tab_ConnectedComponents}, the giant (strongly) connected components of CE1 and CE2 are identical, and in the following only CE1 will be considered.


\section{Adjacency, connectance, asymmetry}

In a first step, we analyzed all considered graphs with respect to basic graph-statistical measures. These include the number of self-loops $N_L$, defined as the number of non-zero diagonal elements $a_{ii}$ in the adjacency matrix, and the total adjacency $A$, defined as the sum over all entries in the adjacency matrix, with diagonal elements (self-loops) counting two:
\begin{equation}
\label{Eq_TotalAdjacency}
A = \sum\limits_{i,j=1}^{N_N} a_{ij} + N_L \, . 
\end{equation}
Using the total adjacency, the number of edges $N_E$ is defined as
\begin{equation}
\label{Eq_NumberOfEdges}
N_E = \left\{ 
	\begin{array}{ll}
		A   & \quad \text{directed} \\
		A/2 & \quad \text{undirected} \, .
	\end{array}
	\right. 
\end{equation}

The asymmetry index $\mathcal{A}$ quantifies the ratio between the number of non-symmetrical edges $N_A$ and symmetrical edges $N_S$, and is given by (\citealp{WassermanFaust94, NewmanEA02, SerranoBoguna03}, but see \citealp{GarlaschelliLoffredo04})
\begin{equation}
\label{Eq_AsymmetryIndex}
\mathcal{A} = \frac{N_A}{A - N_S} \, ,
\end{equation}
where $N_A$ is the number of node pairs $(i, j | j \geq i)$ for which $a_{ij} \neq a_{ji}$, and $N_S$ is the number of node pairs $(i, j | j \geq i)$ for which $a_{ij} = a_{ji} = 1$. It can be shown that $0 \leq \mathcal{A} \leq 1$, and that Eq.~\ref{Eq_AsymmetryIndex} holds for self-looped and non self-looped graphs.

Finally, the graph connectedness (or connectance) $Co$, a measure of relative graph connectivity, is defined as \citep{BoccalettiEA06, Newman10}
\begin{equation}
\label{Eq_Connectedness}
Co 
= \frac{N_E}{N_E^{max}} 
= \left\{ 
	\begin{array}{ll}
		\frac{A}{N_N (N_N+1)} & \quad \text{self-looped} \\
		\frac{A}{N_N (N_N-1)} & \quad \text{not self-looped} ,
	\end{array}
	\right.
\end{equation}
where $N_E^{max}$ denotes the number of possible edges in a complete, i.e. maximally connected, graph:
$$
\label{Eq_NEmax}
N_E^{max} 
= \left\{ 
	\begin{array}{ll}
		N_N (N_N+1) 			 & \quad \text{directed, self-looped} \\
		N_N (N_N-1)  			 & \quad \text{directed, not self-looped} \\ 		\frac{1}{2} N_N (N_N+1)  & \quad \text{undirected, self-looped} \\
		\frac{1}{2} N_N (N_N-1)  & \quad \text{undirected, not self-looped}.
	\end{array}
	\right.
$$
It can be shown that $0 \leq Co \leq 1$. We note that Eq.~\ref{Eq_Connectedness} generalizes the commonly used definition of the connectedness to graphs containing self-loops (e.g., see \citealp{BoccalettiEA06, Newman10}). 

The basic graph-statistical properties of both the directed and symmetrized (undirected) versions of the investigated neural graphs are listed in Table~\ref{Tab_BasicProperties}. Of these graphs, only CE3 has self-loops, accounting for about 0.1\% of the graph's edges, which stem from electrical junctions connecting a node with itself. If not specified otherwise, these self-loops were included in the analysis. 

Naturally, the total adjacency $A$ is larger for the undirected version of the corresponding graphs, as symmetrization of the adjacency matrix only adds edges to a given digraph. By symmetrizing a graph, the total adjacency and, thus, the connectedness can increase by more than 50\%, as in the case of the {\it C. elegans} or macaque brain neural graphs (CE1, CE2: $Co^{ud} \sim 1.828\ Co^{d}$, CE3: $Co^{ud} \sim 1.526\ Co^{d}$, MB1: $Co^{ud} \sim 1.577\ Co^{d}$; $Co^{d}$ and $Co^{ud}$ denote the connectedness of the directed and undirected versions of a graph, respectively). Thus, the consideration of undirected versions of digraphs in the literature might already at this level introduce a significant mischaracterization of investigated graphs.

\begin{figure}[t]
\includegraphics[width=\columnwidth]{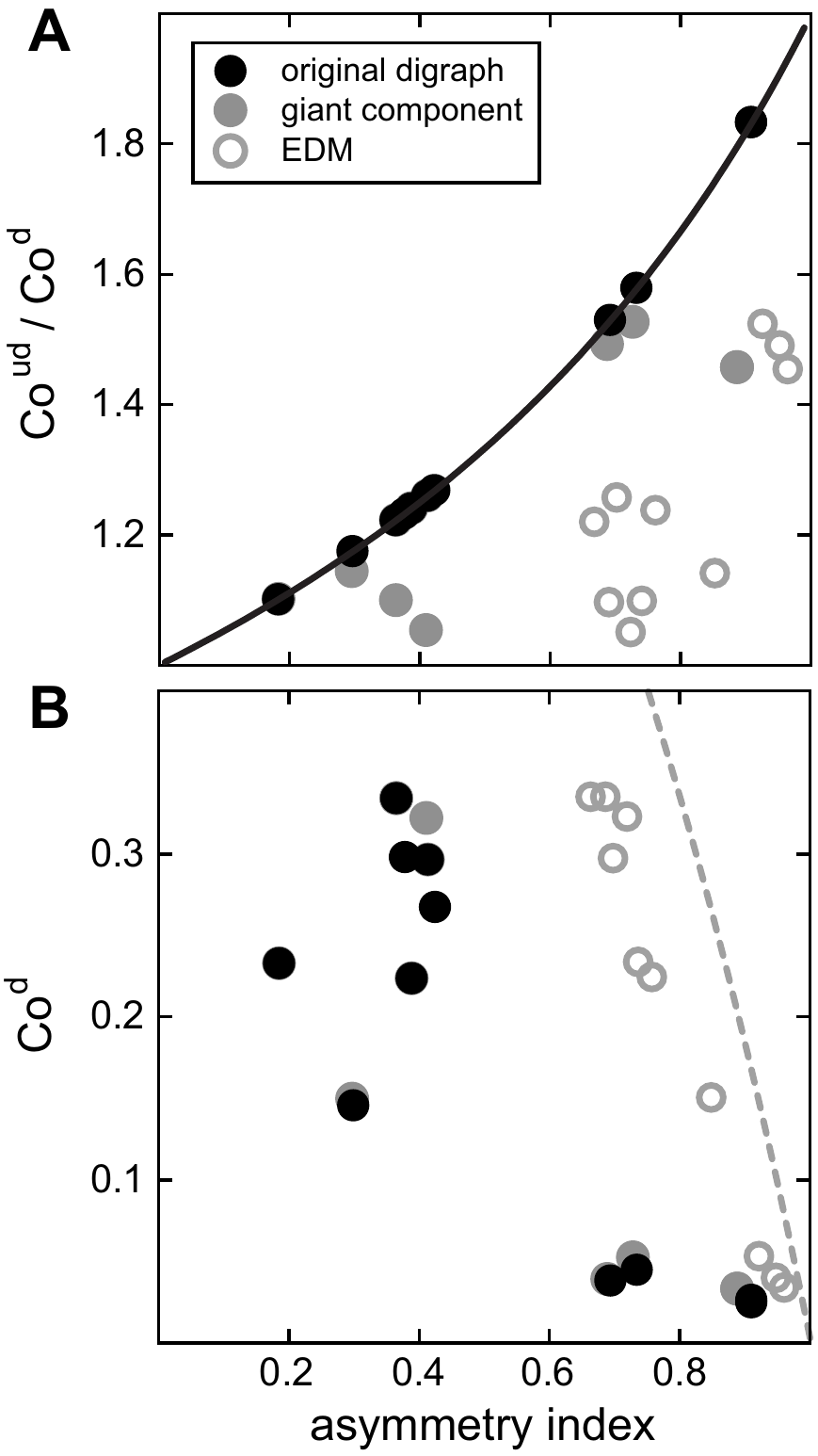}
\caption{\label{Fig_AsymmetryConnectedness}
Relation of connectedness and asymmetry index in various biological neural graphs and their giant connected components. (a): The ratio between connectedness of the symmetrized (undirected) version of a graph ($Co^{ud}$) and its directed original ($Co^{d}$) increases with the asymmetry index of the digraph. The solid line shows the theoretical relation, Eq.~\ref{Eq_RelationACoudCod}. (b): In biological neural digraphs, high connectedness appears to be linked with a lower asymmetry index. The dashed line shows the analytical result for a Erd\H{o}s-R\'enyi graph model, Eq.~\ref{Eq_ConnectednessERmodel}. In both panels, results are shown for the original digraphs (black dots), their giant connected components (grey dots), and corresponding exact degree-matched Erd\H{o}s-R\'enyi graph models (open dots).}
\end{figure}

\begin{table*}[t]
\caption{Connected component analysis of original directed and symmetrized (undirected) versions of various neural graphs. Shown are values for the number of connected components $N_{cc}$, the size of the giant connected component $S_{gcc}$, total adjacency $A$ ($N_E$ given in Eq.~\ref{Eq_NumberOfEdges}), asymmetry index $\mathcal{A}$ ($\mathcal{A} = 0$ for undirected graphs), connectedness $Co$, the minimum and maximum node degree $\delta$ and $\Delta$, respectively ($\delta^{\{in,out\}}$ and $\Delta^{\{in,out\}}$ for directed graphs), and the average node degree $\langle a_i \rangle$ for the giant connected components.}
\label{Tab_ConnectedComponents}
\begin{tabular*}{\textwidth}{l|lllp{9mm}p{9mm}llp{9mm}|llp{9mm}p{9mm}ll}
\hline
     & \multicolumn{8}{c|}{{\bf directed}} & \multicolumn{6}{c}{{\bf undirected}} \\
     & $N_{cc}$ & $S_{gcc}$ & $A$ & $\mathcal{A}$ & $Co$ & $\delta/\Delta^{in}$ & $\delta/\Delta^{out}$ & $\langle a_i \rangle$ & $N_{cc}$ & $S_{gcc}$ & $A$ & $Co$ & $\delta/\Delta$ & $\langle a_i \rangle$ \\
\hline
CC1  &  1 &  95 & 2126 & 0.1829 & 0.2331 &  2/55 &  2/52 & 22.38 &  1 &  95 &  2340 & 0.2566 &  2/61 & 24.63 \\
CC2  &  1 &  52 &  818 & 0.4117 & 0.2968 &  7/32 &  3/34 & 15.73 &  1 &  52 &  1030 & 0.3737 &  7/37 & 19.81 \\
CE1  & 66 & 239 & 1912 & 0.8864 & 0.0333 &  1/42 &  1/38 &  8.00 & 10 & 297 &  4296 & 0.0485 & 1/134 & 14.46 \\
CE2  & 57 & 239 & 1912 & 0.8864 & 0.0333 &  1/42 &  1/38 &  8.00 &  1 & 297 &  4296 & 0.0485 & 1/134 & 14.46 \\
CE3  &  6 & 274 & 2962 & 0.6871 & 0.0393 &  1/82 &  1/57 & 10.80 &  1 & 279 &  4580 & 0.0586 &  2/93 & 16.41 \\
MB1  & 33 & 351 & 6491 & 0.7265 & 0.0525 & 1/103 & 1/108 & 18.49 & 24 & 360 & 10416 & 0.0801 & 1/149 & 28.93 \\
MC1  &  2 &  70 &  745 & 0.2952 & 0.1499 &  2/26 &  2/28 & 10.64 &  1 &  71 &   876 & 0.1714 &  1/28 & 12.34 \\
MC2  & 10 &  85 & 2356 & 0.4092 & 0.3223 &  1/65 &  1/54 & 27.72 &  1 &  94 &  3030 & 0.3393 &  1/74 & 32.23 \\
MNC1 &  1 &  47 &  505 & 0.3866 & 0.2238 &  1/23 &  2/23 & 10.74 &  1 &  47 &   626 & 0.2775 &  3/27 & 13.32 \\
MVC1 &  1 &  30 &  311 & 0.3632 & 0.3344 &  2/19 &  4/20 & 10.37 &  1 &  30 &   380 & 0.4086 &  5/22 & 12.67 \\
MVC2 &  3 &  30 &  311 & 0.3632 & 0.3344 &  2/19 &  4/20 & 10.37 &  1 &  32 &   388 & 0.3674 &  2/22 & 12.13 \\
\hline
\end{tabular*}
\end{table*}

Due to the symmetrization procedure, the connectedness is intrinsically dependent on the asymmetry of the considered graph. The higher the asymmetry $\mathcal{A}$, the more edges will be added, yielding a higher connectedness in the undirected version of a given digraph (Fig.~\ref{Fig_AsymmetryConnectedness}A). A theoretical relation between the ratio of the connectedness for digraphs and their undirected equivalents, and the asymmetry index can be obtained by observing that the total adjacency $A^d = N_A + 2 N_S$ for digraphs takes after symmetrization the form $A^{ud} = 2 (N_A + N_S)$. This yields
$$
A^d = A^{ud} \Big( 1 - \frac{1}{2} \mathcal{A} \Big) \, .
$$
If we assume that the number of nodes in both the directed and its symmetrized version are the same, this gives, together with Eq.~\ref{Eq_Connectedness}, the desired relation
\begin{equation}
\label{Eq_RelationACoudCod}
\frac{Co^{ud}}{Co^d} = \frac{2}{2 - \mathcal{A}} \, ,
\end{equation}
shown in Fig.~\ref{Fig_AsymmetryConnectedness}A (solid). However, the numerical results for the giant component, as well as the corresponding EDM graphs, deviates from Eq.~\ref{Eq_RelationACoudCod} (Fig.~\ref{Fig_AsymmetryConnectedness}A, compare grey and open dots with black solid line). The reason for this deviation is simply that, in the case of undirected graphs, the giant connected components were obtained from their original digraphs after symmetrization (see Methods). This leads to a change in the number of nodes in the constructed giant components of corresponding digraph and undirected graph, therefore Eq.~\ref{Eq_RelationACoudCod} does no longer apply.

Interestingly, a weak relation between connectedness and asymmetry index can also be found when considering digraphs only (Fig.~\ref{Fig_AsymmetryConnectedness}B), with graphs of higher connectedness being associated with a weaker asymmetry. Such a link is expected, however, and can be calculated analytically in the case of directed Erd\H{o}s-R\'enyi (ER) graphs. Excluding (for simplicity) self-loops, the total adjacency is $p N_N (N_N - 1)$, where $p$ denotes the connection probability of a classical ER graph. With the total number of possible edges in a directed, not self-looped ER graph being $N_E^{max} = N_N (N_N - 1)$, the connectedness $Co = p$. Node pairs $(i, j)$ with $a_{ij} = 1 \wedge a_{ji} = 1$ occur here with a probability of $p^2$, and are the only contribution to the number of symmetric edges $N_S$, thus yielding
\begin{equation}
N_S = p^2 \frac{N_E^{max}}{2} \, .
\end{equation}
In a similar fashion, node pairs $(i, j)$ with $a_{ij} = 1 \wedge a_{ji} = 0$ or $a_{ij} = 0 \wedge a_{ji} = 1$ occur with a probability $p(1-p)$ each and contribute to the number of non-symmetrical edges $N_A$, yielding
\begin{equation}
N_A = 2p(1-p) \frac{N_E^{max}}{2} \, .
\end{equation}
With this, the asymmetry index (Eq.~\ref{Eq_AsymmetryIndex}) of a ER graph is given by
\begin{equation}
\label{Eq_AsymmetryERmodel}
\mathcal{A} = 2 \frac{1-Co}{2-Co} \, ,
\end{equation}
which yields the desired relation between connectedness and asymmetry
\begin{equation}
\label{Eq_ConnectednessERmodel}
Co = 2 \frac{\mathcal{A}-1}{\mathcal{A}-2} \, .
\end{equation}

The theoretical result for classical ER graph models (Eq.~\ref{Eq_ConnectednessERmodel}) is independent of the number of nodes, and is shown in Fig.~\ref{Fig_AsymmetryConnectedness}A (dashed line). Although displaying the same qualitative behavior, namely a decrease in the connectedness for increasing asymmetry index, the quantitative results for the investigated neural graphs deviates significantly from the theoretical expectation for ER graph models. Even after the incorporation of the exact node degree distribution using the EDM graph models (Fig.~\ref{Fig_AsymmetryConnectedness}A, grey circles), the results still deviate markedly from that observed in their biological counterparts. This suggests that a random distribution of edges with a given degree distribution cannot account for the relation between connectedness and asymmetry observed in biological neural digraphs. However, adding a correlation between node in- and out-degree will increase the probability of occurrence of node pairs $(i, j)$ with $a_{ij} = a_{ji} = 1$, thus increase the number of symmetric edges $N_S$ and proportionally lower the number of non-symmetric edges $N_A$. According to Eq.~\ref{Eq_AsymmetryIndex}, this will effectively lead to a a decrease in $\mathcal{A}$ for a given connectedness. As we will show below, such a correlation between a node's in- and out-degree is indeed what we observe in the investigated biological graphs.


\section{Node degrees}

To further characterize structural aspects of biological neural digraphs, we calculated the node in- and out-degrees
\begin{eqnarray}
a_i^{in} & = & \sum\limits_{j=1}^{N_N} a_{ji} \label{Eq_NodeInDegree} \\ 
a_i^{out} & = & \sum\limits_{j=1}^{N_N} a_{ij} \label{Eq_NodeOutDegree}
\end{eqnarray}
as well as the node degree 
\begin{equation}
\label{Eq_NodeDegree}
a_i = \sum\limits_{j=1}^{N_N} a_{ij} + a_{ii}  
\end{equation}
for the corresponding symmetrized graphs. Note that the definition of the node degree in Eq.~\ref{Eq_NodeDegree} deviates from that commonly employed through the inclusion of self-loops, which are considered as contributing two edges to adjacency relations (one in-edge and one out-edge pointing to the same node, see above). This definition is more natural, as it is a direct result of the definition of $a_i$ for digraphs in Eqs.~\ref{Eq_NodeInDegree} and \ref{Eq_NodeOutDegree}, and by defining terms in this way, undirected graphs become a special case of digraphs, which then assume the more fundamental role. With this, the handshaking lemma, which provides a consistency relation linking the sum over all node degrees with the total adjacency of a graph, takes a more general form valid for both directed and undirected self-looped graphs:
\begin{equation}
\label{Eq_HandshakingLemma}
\sum\limits_{i=1}^{N_N} a_i = 2 ( A - N_L ) \, ,
\end{equation}
where $a_i = a_i^{in} + a_i^{out}$ for digraphs.

Given the node (in/out-) degrees of a graph, we define the minimum and maximum node (in/out-) degree, respectively, as $\delta$ ($\delta^{\alpha}$) and $\Delta$ ($\Delta^{\alpha}$), $\alpha \in \{ in, out \}$. Furthermore, the average node in/out-degree $\langle a_i^{\alpha} \rangle$ for directed and average node degree $\langle a_i \rangle$ for undirected graphs is given by
\begin{equation}
\label{Eq_AverageNodeDegree}
\langle a_i \rangle
= \left\{ 
	\begin{array}{ll}
		\frac{1}{N_N} \sum\limits_{i=1}^{N_N} a_i^{in} & = \langle a_i^{in} \rangle = \langle a_i^{out} \rangle = \frac{1}{N_N} \sum\limits_{i=1}^{N_N} a_i^{out} \\
        & \quad \text{directed} \\
		\frac{1}{N_N} \sum\limits_{i=1}^{N_N} a_i & \quad \text{undirected} .
	\end{array}
	\right.
\end{equation}
Note that due to the handshaking lemma, Eq.~\ref{Eq_HandshakingLemma}, we have $\langle a_i^{in} \rangle = \langle a_i^{out} \rangle$.

\begin{table}[b]
\caption{\label{Tab_NodeDegreePDFfits1}
Power law fits of the tail degree distributions for various biological neural graphs. The values give the best fitting parameters $\alpha^{\{in,out\}}$ and $\alpha$ according to Eq.~\ref{Eq_PowerLaw} for the node in/out-degree and node degree PDFs of the directed and undirected versions of the graphs, respectively. Values in parentheses are excluded from Fig.~\ref{Fig_NodeDegreePDFfits}.}
\begin{tabular*}{\columnwidth}{l|p{2mm}p{16mm}p{16mm}l}
\hline
& \multicolumn{4}{c}{\bf{power-law model}} \\
& 
& $\alpha^{in}$
& $\alpha^{out}$ 
& $\alpha$ \\
\hline
CC1  & & 3.4448 & 2.9610 & 3.8817 \\
CC2  & & 3.4922 & 3.7824 & 2.3411 \\
CE1  & & 2.8122 & 2.8068 & 3.1565 \\
CE3  & & 2.7072 & 3.4023 & 2.7054 \\
MB1  & & 2.1281 & 2.2697 & 2.3336 \\
MC1  & & 3.0592 & 2.7606 & 2.6497 \\
MC2  & & 4.0459 &(6.6290)& 3.6015 \\
MNC1 & & 1.7702 & 1.7275 & 2.0251 \\
MVC1 & & 1.9798 & 2.7790 & 2.7924 \\
MVC2 & & 1.7711 & 2.7790 & 2.7573 \\
\hline
\end{tabular*}
\end{table}

Results for the investigated biological graphs are summarized in Table~\ref{Tab_NodeDegreeProperties}. In two of the investigated graphs (CE1 and MB1) the minimal node degrees $\delta^{\alpha}$ and $\delta$ in both the directed and undirected version, respectively, are zero. Furthermore, the minimal total node degree in these graphs is also zero, indicating the existence of nodes without edges. Further analysis revealed that in the directed version of CE1, the number of weakly connected components, i.e. subgraphs whose nodes are connected by at least one directed edge to other nodes in the same subgraph, is 10, with the size of the largest weakly connected component being 297 nodes. Thus, with a total of 306 in this graph, the remaining 9 components share 9 nodes, i.e. each of the remaining weakly connected components contains only one isolated node. The same argumentation applies to the undirected version of CE1. The MB1 graph has one giant weakly connected component with 351 nodes, and the remaining 32 connected components share 32 isolated nodes. The existence of these isolated nodes in the neural graphs suggests that the mapping of these graphs is incomplete, as such nodes are very unlikely to have a functional or structural meaning. Therefore, in the remainder of this study, we will focus our analysis on the giant (strongly) connected component of each investigated graph (see Methods -- Connected components).


\section{Node degree distributions}

A prevalent theme in the literature of the past two decades is scale-free properties of various real-world systems, typically investigated by fitting corresponding physical quantities with power-law distributions. However, recently it was pointed out that in many cases such a fit provides only a poor description of the true behavior, or at best a faithful representation in only a narrow region of the investigated quantities' value range \citep{ClausetEA09, LimaMendezVanHelden09}. This is especially crucial when considering small systems, for which boundary effects cannot feasibly be neglected. Moreover, claims of scale-free properties, with little or no support from experimental data, may distract further search for mechanisms by which such networks form and develop.

\begin{table*}[t]
\caption{\label{Tab_NodeDegreePDFfits2}
Power law with cutoff and gamma fits of the node degree distributions for various biological neural graphs. The values give the best fitting parameters $\alpha^{\{in,out\}}$, $\lambda^{\{in,out\}}$ and $\alpha$, $\lambda$ according to Eq.~\ref{Eq_PowerLawWithCutoff} as well as $\theta^{\{in,out\}}$, $k^{\{in,out\}}$ and $\theta$, $k$ according to Eq.~\ref{Eq_Gamma} for the node in/out-degree and node degree PDFs of the directed and undirected versions of the graphs, respectively. Both fitting models are qualitatively equivalent. Values in parentheses are excluded from Fig.~\ref{Fig_NodeDegreePDFfits}.}
\begin{tabular*}{\textwidth}{l|lp{7mm}lp{7mm}lp{8mm}|lp{8mm}lp{9mm}p{8mm}l}
\hline
& \multicolumn{6}{c|}{\bf{power-law with cutoff model}}
& \multicolumn{6}{c}{\bf{gamma model}} \\
& $\alpha^{in}$ & $\lambda^{in}$
& $\alpha^{out}$ & $\lambda^{out}$
& $\alpha$ & $\lambda$
& $\theta^{in}$ & $k^{in}$
& $\theta^{out}$ & $k^{out}$
& $\theta$ & $k$ \\
\hline
CC1  & -1.9885 & 0.1212 &  -1.4334 & 0.0995 &  -1.4159 & 0.0885 &  8.2456 & 2.9891 & 10.0439 &  2.4338 & 11.3001 &  2.4163 \\
CC2  & -5.4192 & 0.4135 &  -1.8189 & 0.1680 &  -3.6622 & 0.2293 &  2.4169 & 6.4233 &  5.9539 &  2.8193 &  4.3598 &  4.6636 \\
CE1  & -0.7030 & 0.2277 &  -0.1054 & 0.1333 &  -2.1114 & 0.2281 &  4.1422 & 1.7442 &  6.3163 &  1.1787 &  4.3814 &  3.1114 \\
CE3  & -1.6139 & 0.2769 &  -1.1791 & 0.2084 &  -2.7632 & 0.2612 &  3.5851 & 2.6229 &  4.7285 &  2.1953 &  3.8276 &  3.7635 \\
MB1  &  0.3211 & 0.0351 &   0.3322 & 0.0281 &  -0.1318 & 0.0361 & 21.8775 & 0.7280 & 28.9742 &  0.7063 & 26.8478 &  1.1439 \\
MC1  & -1.6364 & 0.2462 &  -1.2988 & 0.2077 &  -1.3638 & 0.1684 &  4.0623 & 2.6365 &  4.8144 &  2.2989 &  5.8766 &  2.3823 \\
MC2  & -5.4764 & 0.1988 &(-20.7122)& 0.6916 &(-15.7716)& 0.4560 &  5.0266 & 6.4798 &  1.4455 &(21.7190)&  2.1922 &(16.7780)\\
MNC1 & -0.6411 & 0.1180 &  -1.7654 & 0.2445 &  -1.7342 & 0.1913 & -6.4050 & 1.6074 &  4.0904 &  2.7656 &  5.2265 &  2.7344 \\
MVC1 & -4.4855 & 0.5045 &  -3.4864 & 0.4100 &  -7.7057 & 0.6615 &  1.9820 & 5.4858 &  2.4385 &  4.4869 &  1.5100 &  8.7159 \\
MVC2 & -4.4855 & 0.5045 &  -3.4864 & 0.4100 &  -6.1531 & 0.5291 &  1.9820 & 5.4858 &  2.4385 &  4.4869 &  1.8898 &  7.1533 \\
\hline
\end{tabular*}
\end{table*}

\begin{table*}[t]
\caption{\label{Tab_NodeDegreePDFfits3}
Stretched exponential and log-normal fits of the node degree distributions for various biological neural graphs. The values give the best fitting parameters $\beta^{\{in,out\}}$, $\lambda^{\{in,out\}}$ and $\beta$, $\lambda$ according to Eq.~\ref{Eq_StretchedExponential} as well as $\mu^{\{in,out\}}$, $\sigma^{\{in,out\}}$ and $\mu$, $\sigma$ according to Eq.~\ref{Eq_LogNormal} for the node in/out-degree and node degree PDFs of the directed and undirected versions of the graphs, respectively. Values in parentheses are excluded from Fig.~\ref{Fig_NodeDegreePDFfits}.}
\begin{tabular*}{\textwidth}{l|p{7mm}p{15mm}p{8mm}p{14mm}p{7mm}l|p{7mm}p{7mm}p{7mm}p{7mm}p{7mm}l}
\hline
& \multicolumn{6}{c|}{\bf{stretched exponential model}}
& \multicolumn{6}{c}{\bf{log-normal model}} \\
& $\beta^{in}$ & $\lambda^{in}$ 
& $\beta^{out}$ & $\lambda^{out}$
& $\beta$ & $\lambda$
& $\mu^{in}$ & $\sigma^{in}$
& $\mu^{out}$ & $\sigma^{out}$
& $\mu$ & $\sigma$ \\
\hline
CC1   & 2.0180 & 1.4140$\cdot 10^{-3}$ & 1.7297 & 3.5818$\cdot 10^{-3}$ & 1.7717 & 2.6172$\cdot 10^{-3}$ & 3.1299 & 0.6495 & 3.0851 & 0.7146 & 3.2057 & 0.7395 \\
CC2   & 2.8725 & 3.3298$\cdot 10^{-4}$ & 1.8293 & 4.9790$\cdot 10^{-3}$ & 2.3767 & 6.7242$\cdot 10^{-4}$ & 2.7072 & 0.4171 & 2.7145 & 0.6526 & 2.9535 & 0.4941 \\
CE1   & 1.3953 & 5.6656$\cdot 10^{-2}$ & 1.0540 & 1.0567$\cdot 10^{-1}$ & 1.9149 & 6.1237$\cdot 10^{-3}$ & 1.8429 & 0.9189 & 1.8158 & 1.1163 & 2.5433 & 0.5951 \\
CE3   & 1.8187 & 1.5470$\cdot 10^{-2}$ & 1.6044 & 2.0917$\cdot 10^{-2}$ & 2.1206 & 3.0620$\cdot 10^{-3}$ & 2.1561 & 0.7096 & 2.2170 & 0.7773 & 2.6015 & 0.5483 \\
MB1   & 0.7751 & 1.0642$\cdot 10^{-1}$ & 0.7816 & 8.9299$\cdot 10^{-2}$ & 1.1078 & 2.1859$\cdot 10^{-2}$ & 2.5100 & 1.5886 & 2.7913 & 1.8178 & 3.2205 & 1.3011 \\
MC1   & 1.7632 & 1.3072$\cdot 10^{-2}$ & 1.6654 & 1.6043$\cdot 10^{-2}$ & 1.7744 & 8.4554$\cdot 10^{-3}$ & 2.2461 & 0.6905 & 2.2642 & 0.7655 & 2.5209 & 0.7640 \\
MC2   & 3.1522 &(1.6068$\cdot 10^{-5}$)&(5.0710)&(2.2765$\cdot 10^{-8}$)&(4.6665)&(4.3361$\cdot 10^{-8}$)& 3.4657 & 0.4108 & 3.4417 & 0.2109 & 3.5959 & 0.2497 \\
MNC1  & 1.4838 & 1.9558$\cdot 10^{-2}$ & 1.8212 & 1.0289$\cdot 10^{-2}$ & 1.8120 & 7.0028$\cdot 10^{-3}$ & 2.5110 & 0.8737 & 2.3030 & 0.6657 & 2.5353 & 0.6771 \\
MVC1  & 2.5811 & 1.8229$\cdot 10^{-3}$ & 2.3898 & 2.8284$\cdot 10^{-3}$ & 3.4395 & 1.1885$\cdot 10^{-4}$ & 2.3429 & 0.4509 & 2.3330 & 0.5116 & 2.5502 & 0.3503 \\
MVC2  & 2.5811 & 1.8229$\cdot 10^{-3}$ & 2.3898 & 2.8284$\cdot 10^{-3}$ & 3.0058 & 3.4508$\cdot 10^{-4}$ & 2.3429 & 0.4509 & 2.3330 & 0.5116 & 2.5726 & 0.3887 \\
\hline
\end{tabular*}
\end{table*}

In order to assess the extent to which the power-law provides a valid description of structural characteristics of biological neural graphs, we studied the node degree probability density functions (PDFs; node in-degree and out-degree PDFs for digraphs) of these graphs. The power law model is defined by
\begin{equation}
\label{Eq_PowerLaw}
\rho^{pl}(a; \alpha) = a_{min}^{\alpha-1} (\alpha-1) a^{-\alpha} \, ,
\end{equation}
where $a$ denotes the node degree and $a_{min}$ the lower bound of the fitting interval. In addition, we applied other fitting models proposed in the literature (see \citealp{ClausetEA09}). The second model considered was the ``power-law with cutoff'', defined by
\begin{equation}
\label{Eq_PowerLawWithCutoff}
\rho^{plwc}(a; \alpha, \lambda) = \frac{1}{\Gamma[1-\alpha, \lambda a_{min}]} \lambda^{1-\alpha} a^{-\alpha} e^{-\lambda a} \, ,
\end{equation}
where $\Gamma[s,x]$ is the incomplete Gamma function. We note that due to the exponential term, this model carries none of the implications commonly associated with the power-law model (i.e. scale-free characteristics), as this term replaces the power-law properties at both tails of the distribution, leaving only the dominance of the power-law behavior within a certain charactistic scale. However, to avoid confusion and to remain in accordance with the literature \citep{ClausetEA09}, we retain this terminology through the remainder of the paper.

\begin{figure*}[t]
\includegraphics[width=\textwidth]{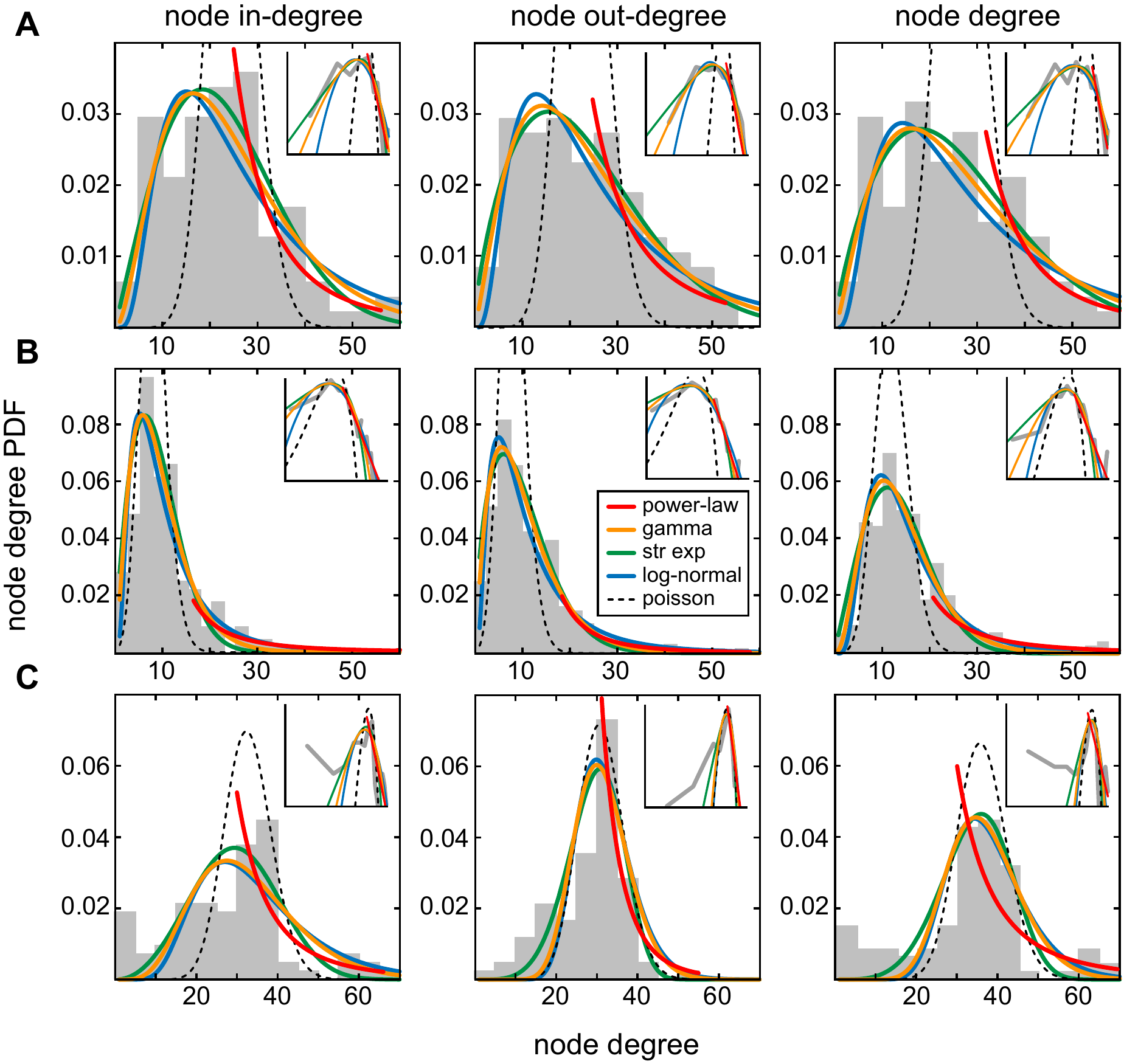}
\caption{\label{Fig_NodeDegreePDFs}
Representative examples of node degree probability density functions and their fits. (a): CC1, (b): CE3, (c): MC2. For digraphs, the PDFs (grey bar plots) of node in-degree (left) and node out-degree (middle) are shown, for 
undirected graphs PDFs of the node degree (right). Best fits of the node degree distributions are provided by the power-law with cutoff / gamma model (Eq.~\ref{Eq_PowerLawWithCutoff}; orange solid), followed by the stretched exponential model (Eq.~\ref{Eq_StretchedExponential}; green solid) and log-normal model (Eq.~\ref{Eq_LogNormal}; blue solid). The power law model (Eq.~\ref{Eq_PowerLaw}; red solid) provides an approximate fit of the tail of a given node degree PDF only. In all considered graphs, the Poisson model (Eq.~\ref{Eq_Poisson}; black dashed) did not deliver an acceptable fit of the data. Insets show the corresponding data in log-log representation.}
\end{figure*}

\begin{figure*}[t]
\includegraphics[width=\textwidth]{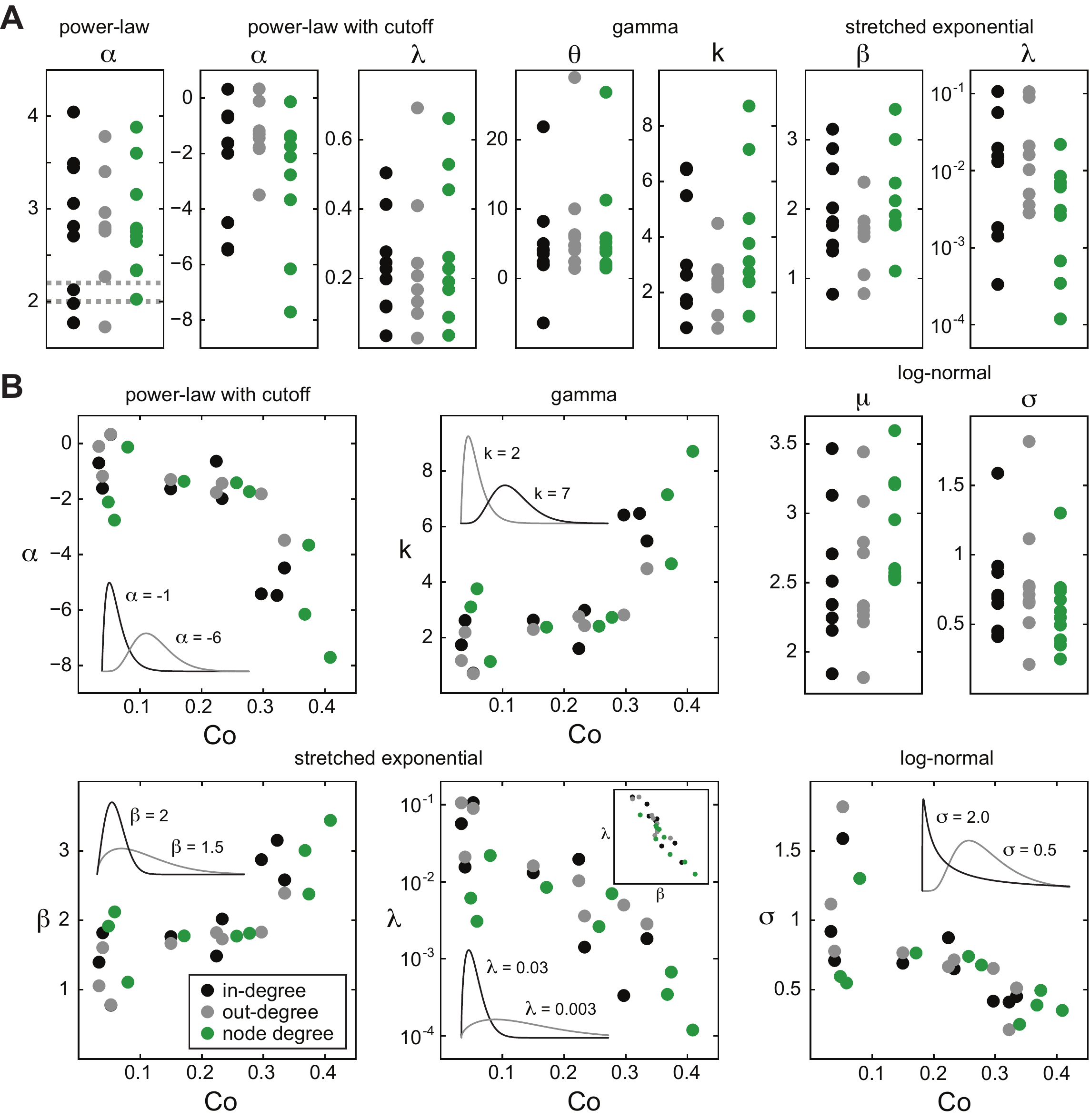}
\caption{\label{Fig_NodeDegreePDFfits}
Best parameter fits for node degree PDFs of various biological neural graphs.
(a): Distribution of best fitting parameters for various models (Eqs.~\ref{Eq_PowerLaw} to \ref{Eq_LogNormal}). (b): For some models, a correlation between fitted model parameters and graph connectedness $Co$ was found. The insets show the qualitative role of the parameter on the model distribution.}
\end{figure*}

Further models utilized were the stretched exponential model
\begin{equation}
\label{Eq_StretchedExponential}
\rho^{se}(a; \beta, \lambda) = \beta \lambda e^{\lambda a_{min}^{\beta}} a^{\beta-1} e^{-\lambda a^{\beta}} 
\end{equation}
and gamma model
\begin{equation}
\label{Eq_Gamma}
\rho^{g}(a; \theta, k) = \frac{1}{\Gamma[k]} a^{k-1} e^{-a/\theta} \theta^{-k} \, .
\end{equation}
The latter is equivalent to the power-law with cutoff model when considering $k \leftrightarrow (1-\alpha)$ and $\theta \leftrightarrow 1/\lambda$. Finally, the log-normal model
\begin{equation}
\label{Eq_LogNormal}
\rho^{ln}(a; \mu, \sigma) = N \frac{1}{a} \exp \left[-\frac{(\ln a - \mu)^2}{2 \sigma^2}\right]
\end{equation}
with
$$
N = \sqrt{\frac{2}{\pi \sigma^2}} \left( \mbox{erfc}\left[ \frac{\ln a_{min} - \mu}{\sqrt{2 \sigma^2}} \right]\right)^{-1}
$$
and the Poisson model
\begin{equation}
\label{Eq_Poisson}
\rho^{p}(a; \mu) = \tilde{N} \frac{1}{\Gamma[a+1]} \mu^{a}
\end{equation}
with 
$$
\tilde{N} = \left( e^{\mu} - \sum\limits_{k=0}^{a_{min}-1} \frac{\mu^k}{\Gamma[k+1]} \right)^{-1}
$$
were considered. In all cases, $a_{min} = 1$ was used, except for the power-law model, which only allowed fitting the tail of the degree distributions. Moreover, all fitted models were constrained by the normalization condition
\begin{equation}
\int\limits_{a_{min}}^{\infty} \rho^{\alpha}(a; \bullet) \, da = 1 \, ,
\end{equation}
where $\alpha \in \{ plwc, se, g, ln, p \}$ and $\bullet$ stands for the set of parameters of the corresponding model, with the exception of the power-law model. As the latter fits only the tail of a given PDF, the normalization constant was adjusted to the fraction of the node degree PDF $\rho(a_i)$ above the lower bound, i.e.
\begin{equation}
\int\limits_{a_{min}}^{\infty} \rho^{pl}(a_i; \alpha) = \int\limits_{a_{min}}^{\infty} \rho(a_i) \, .
\end{equation}

Representative examples of the node degree PDFs for the cat cortex graph (CC1) and the neural connectivity graph of {\it C. elegans} (CE3) are shown in Fig.~\ref{Fig_NodeDegreePDFs}A and Fig.~\ref{Fig_NodeDegreePDFs}B, respectively. Among the graphs considered, only the node degree distributions of MC2 did not allow for a reasonable fit with any of the above models, both in the directed and undirected version (Fig.~\ref{Fig_NodeDegreePDFs}C). This hints either at a very peculiar connectivity pattern in this graph, as it describes the cortical connectivity pattern in only one hemisphere, or its incomplete representation due to missing experimental data. Additionally, in no case the Poisson model (Eq.~\ref{Eq_Poisson}) delivered an acceptable fit of the node degree PDFs (Fig.~\ref{Fig_NodeDegreePDFs}, black dashed), for which reason it was excluded from further consideration.

The obtained best fitting parameters of the node degree models, using the nonlinear least-squares method in {\it Mathematica}, are summarized in Tables~\ref{Tab_NodeDegreePDFfits1} to \ref{Tab_NodeDegreePDFfits3}, and visualized in Fig.~\ref{Fig_NodeDegreePDFfits}A. A detailed presentation of the fitting results using the various models, including the standard error, t-statistics and P-value of the fitted parameters, as well an analysis of the decomposition of the variation in the data attributable to the fitted function and to the residual errors (ANOVA test) along with the root-mean-square difference between actual and predicted values can be found in the Supplemental Material, Fitting of node-degree distributions.

Interestingly, in all fitted models, except the power-law, a weak correlation between fitted parameters and the graph connectedness $Co$ for both node in/out-degree (digraphs) and node degree (undirected graphs) PDFs was found (Fig.~\ref{Fig_NodeDegreePDFfits}B). Specifically, the $\alpha$ parameter in the power-law with cutoff model appears to decrease with increasing connectedness (Fig.~\ref{Fig_NodeDegreePDFfits}B, top left). This may reflect the fact that smaller $\alpha$ values lead to broader distributions, which are expected for more strongly connected graphs. Due to the relation $k = 1 - \alpha$ between $\alpha$ of the power-law with cutoff model and gamma model, a similar relation holds for gamma fits of the PDFs, with $k$ increasing for higher $Co$ (Fig.~\ref{Fig_NodeDegreePDFfits}B, top right). 

\begin{figure*}[t]
\includegraphics[width=\textwidth]{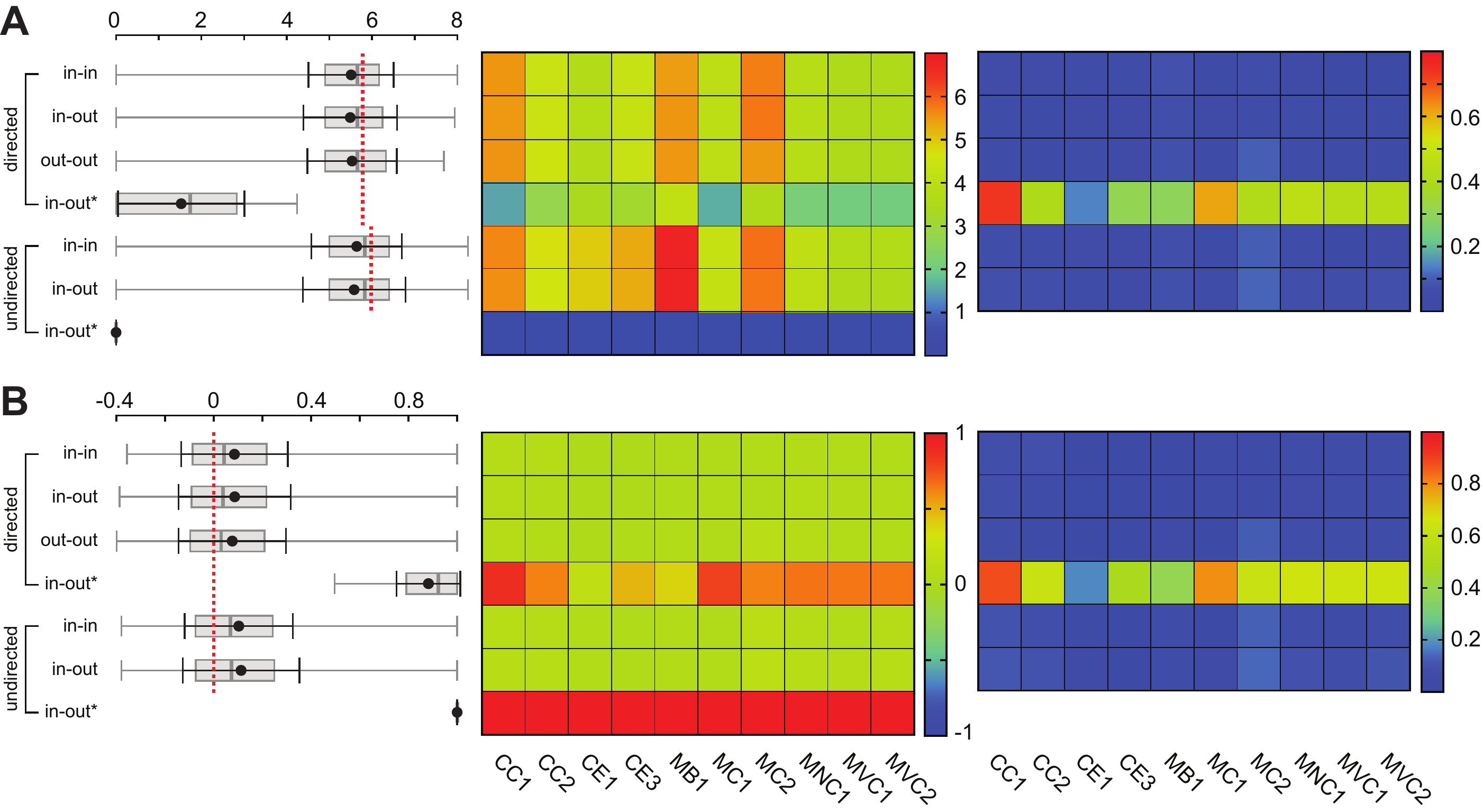}
\caption{\label{Fig_EuclideanDistances_Correlations}
Structural equivalence analysis of neural graphs. A: Euclidean distance of node adjacencies for biological neural graphs. Left: Representative box plots (grey), mean and standard deviation (black) of Euclidean distance measures, Eqs.~\ref{Eq_EuclideanDistanceInIn}-\ref{Eq_EuclideanDistanceOutOut} ({\it in-in}: $D_{ij, i \neq j}^{in-in}$, {\it in-out}: $D_{ij}^{in-out}$, {\it out-out}: $D_{ij, i \neq j}^{out-out}$, {\it in-out*}: $D_{ii}^{in-out}$), for directed and undirected versions of CC1. The red line indicates the value expected for corresponding Erd\H{o}s-R\'enyi graph models. Middle: Mean of Euclidean distance of node adjacencies for all investigated biological neural graphs. Right: relative deviation $\Delta^{ER} D^{\alpha-\beta}$ (Eq.~\ref{Eq_EuclideanDistanceDelta}) from corresponding Erd\H{o}s-R\'enyi graph models. B: Correlation coefficient of node adjacencies for biological neural graphs. Left: Representative box plots (grey), mean and standard deviation (black) of correlation coefficients, Eqs.~\ref{Eq_CorrelationCoefficientInIn}-\ref{Eq_CorrelationCoefficientOutOut} ({\it in-in}: $R_{ij, i \neq j}^{in-in}$, {\it in-out}: $R_{ij}^{in-out}$, {\it out-out}: $R_{ij, i \neq j}^{out-out}$, {\it in-out*}: $R_{ii}^{in-out}$), for directed and undirected versions of CC1. The red line indicates the value expected for corresponding Erd\H{o}s-R\'enyi graph models. Middle: Mean of the correlation coefficient of node adjacencies for all investigated biological neural graphs. Right: Relative deviation $\left| \overline{R_{ij}^{\alpha-\beta}} - \overline{R_{ij}^{ER\ \alpha-\beta}} \right|$ from corresponding Erd\H{o}s-R\'enyi graphs ($\overline{R_{ij}^{ER\ \alpha-\beta}} = 0$, $\alpha, \beta \in \{ in, out \}$).}
\end{figure*}

Both parameters of the stretched exponential model, $\beta$ and $\lambda$ show a weak dependency on the connectedness of the graphs, with $\beta$ increasing and $\lambda$ decreasing for increasing $Co$ (Fig.~\ref{Fig_NodeDegreePDFfits}B, bottom left and middle). For $\lambda$, this dependency reflects again the expected broader distribution of node degrees for stronger connected graphs, as smaller $\lambda$ yields a broader distribution in the stretched exponential model. Interestingly, for the $\beta$ parameter in this model this relation appears to be inverted, with smaller $\beta$ values leading to broader distributions but being associated with smaller connectedness. However, the stretched exponential model displays a strong correlation between the fitted parameters (Fig.~\ref{Fig_NodeDegreePDFfits}B, bottom middle, inset), which is not the case in the other 2-parameter models. Thus, the impact of both $\beta$ and $\lambda$ on the shape of the distribution cannot be considered as independent, and explains the peculiar behavior when both parameters are considered to be independent. Finally, a weak correlation between connectedness and $\sigma$ of the log-normal model was found, with broader distributions (smaller $\sigma$) being associated with higher connectedness.

For assessing the quality of the different models, we compared the root-mean-squares of the fit residuals, i.e. differences between the actual and predicted node degree values. We found that the power-law with cutoff and gamma model provided, on average, the best fits, closely followed by the stretched exponential and log-normal model. This evaluation is consistent with the conclusion reached in \citet{ClausetEA09}. Most interestingly, the nature of these node degree distributions (gamma or power law with cutoff) could be consistent with a simple local mechanism responsible for generating neural graphs. In this way, we may conceive of a graph generation mechanism more parsimonious than those currently in the literature, such as preferential attachment (first discussed as the "Matthew effect" in \citealp{Merton68}; see also \citealp{BarabasiAlbert99}), in which nonlocal knowledge of the degree distribution is required on the level of each individual node (see Discussion).


\section{Structural equivalence}

In order to assess the similarity of the connectivity pattern of individual nodes, various measures of structural equivalence were defined and used in the literature. Here, two nodes are defined as structurally equivalent if they share the same pattern of relationships with all other nodes in a given graph. A first coarse measure quantifying a pattern of relationships among nodes in digraphs is the Euclidean distance between rows and columns of the adjacency matrix \citep{BoccalettiEA06}, defined as
\begin{align}
D_{ij}^{in-in} & = \left\{ \sum\limits_{k=1}^{N_N} ( a_{ki} - a_{kj} )^2 \right\}^{1/2} 
\label{Eq_EuclideanDistanceInIn} \displaybreak[3] \\
D_{ij}^{in-out} & = \left\{ \sum\limits_{k=1}^{N_N} ( a_{ki} - a_{jk} )^2 \right\}^{1/2}
\label{Eq_EuclideanDistanceInOut} \displaybreak[3] \\
D_{ij}^{out-out} & = \left\{ \sum\limits_{k=1}^{N_N} ( a_{ik} - a_{jk} )^2 \right\}^{1/2} 
\label{Eq_EuclideanDistanceOutOut} \, .
\end{align}
Note that here $[D_{ij}^{in-out}]^T = D_{ji}^{in-out} = D_{ij}^{out-in}$, leaving the three independent measures of Euclidean distance in Eqs.~\ref{Eq_EuclideanDistanceInIn}-\ref{Eq_EuclideanDistanceOutOut}. The above definition holds for digraphs with self-loops. If self-loops are excluded, the sum in Eqs.~\ref{Eq_EuclideanDistanceInIn}-\ref{Eq_EuclideanDistanceOutOut} runs over $k \neq \{i, j\}$. The above definitions hold for undirected graphs as well. However, due to the symmetry of the adjacency matrix in this case, we have in addition the relation $D_{ij}^{in-in} = D_{ij}^{out-out} = D_{ij}^{in-out}$, thus leaving only one independent Euclidean distance measure. Moreover, in the case of undirected graphs, $D_{ii}^{in-out} = 0$, which reflects the fact that here, for each given node, the columns and rows of the adjacency matrix are identical. 

According to the notion of structural equivalence, two structurally perfectly equivalent nodes will have identical entries in their corresponding rows and columns in the adjacency matrix. With Eqs.~\ref{Eq_EuclideanDistanceInIn} and \ref{Eq_EuclideanDistanceOutOut}, one thus expects an Euclidean distance $D_{ij}^{in-in} = D_{ij}^{out-out} = 0$. A similar conclusion can, however, not be made for  $D_{ij}^{in-out}$, as the notion of perfect structural equivalence between two nodes does not require a matching pattern in the incoming connection of one node and outgoing connection of another node. 

We calculated the Euclidean distance of node adjacencies, Eqs.~\ref{Eq_EuclideanDistanceInIn}-\ref{Eq_EuclideanDistanceOutOut}, for the giant connected component of the investigated biological neural graphs. To statistically evaluate the distance between two nodes, we considered various subsets of the obtained $N_N \times N_N$ matrices $D_{ij}^{\alpha-\beta}$, $\alpha, \beta \in \{ in, out \}$. First, $D_{ij}^{in-in}, i \neq j$ provides the Euclidean distances between the incoming edges of two different nodes. Secondly, $D_{ij}^{in-out} = D_{ji}^{out-in}$ provides the Euclidean distance between incoming and outgoing edges of two nodes, including the same node. $D_{ij}^{out-out}, i \neq j$ yields the distances between outgoing edges of two different nodes. Finally, $D_{ii}^{in-out}$ contains the Euclidean distances between incoming and outgoing edges of the same node. For each of these subsets of Euclidean distances, we calculated the mean, standard deviation, minimum and maximum value, first and third quartile and median. 

Representative examples of Euclidean distance PDFs and their statistical analysis are shown in Fig.~\ref{Fig_EuclideanDistances_Correlations}A (left; a complete representation of Euclidean distances can be found in the Supplemental Material, Data Tables). We found that in digraphs the mean and median of $D_{ij, i \neq j}^{in-in}$, $D_{ij}^{in-out}$ and $D_{ij, i \neq j}^{out-out}$, and in undirected graphs the mean and median of $D_{ij, i \neq j}^{in-in}$ and $D_{ij}^{in-out}$ are almost identical, a behavior expected from a random, i.e. independent, distribution of edges in different nodes. This behavior was shared among all investigated graphs (Fig.~\ref{Fig_EuclideanDistances_Correlations}A, middle and right). 

In not self-looped random graphs with connectedness $Co$, $a_{ij} = 1$ with probability $Co$ (see above). Thus, two adjacencies with $a_{ij} = 1$ and $a_{mn} = 0$ will occur with probability $Co (1-Co)$. As the latter adjacency relations constitute the only contributions to the Euclidean distance, 
\begin{equation}
\label{Eq_EuclideanDistanceER}
D_{ij}^{\alpha-\beta} = \sqrt{2 N_N Co (1 - Co)} \, ,
\end{equation} 
$\alpha, \beta \in \{ in, out \}$, for Erd\H{o}s-R\'enyi graphs (Fig.~\ref{Fig_EuclideanDistances_Correlations}A, left, red dotted). The relative deviation
\begin{equation}
\label{Eq_EuclideanDistanceDelta}
\Delta^{ER} D^{\alpha-\beta} = \frac{\left| \overline{D_{ij}^{\alpha-\beta}} - \overline{D_{ij}^{ER\ \alpha-\beta}} \right|}{\overline{D_{ij}^{ER\ \alpha-\beta}}} \, ,
\end{equation}
where $\overline{D_{ij}^{\alpha-\beta}}$ and $\overline{D_{ij}^{ER\ \alpha-\beta}}$ denote the mean of the Euclidean distances $D_{ij}^{\alpha-\beta}$ for a given graph and its corresponding Erd\H{o}s-R\'enyi graph, was found to be almost zero (Fig.~\ref{Fig_EuclideanDistances_Correlations}A, right, blue end of color range). This observation provides further evidence suggesting that the distribution of edges in different nodes of biological neural graphs follows a simple random pattern.

\begin{figure}[t]
\includegraphics[width=\columnwidth]{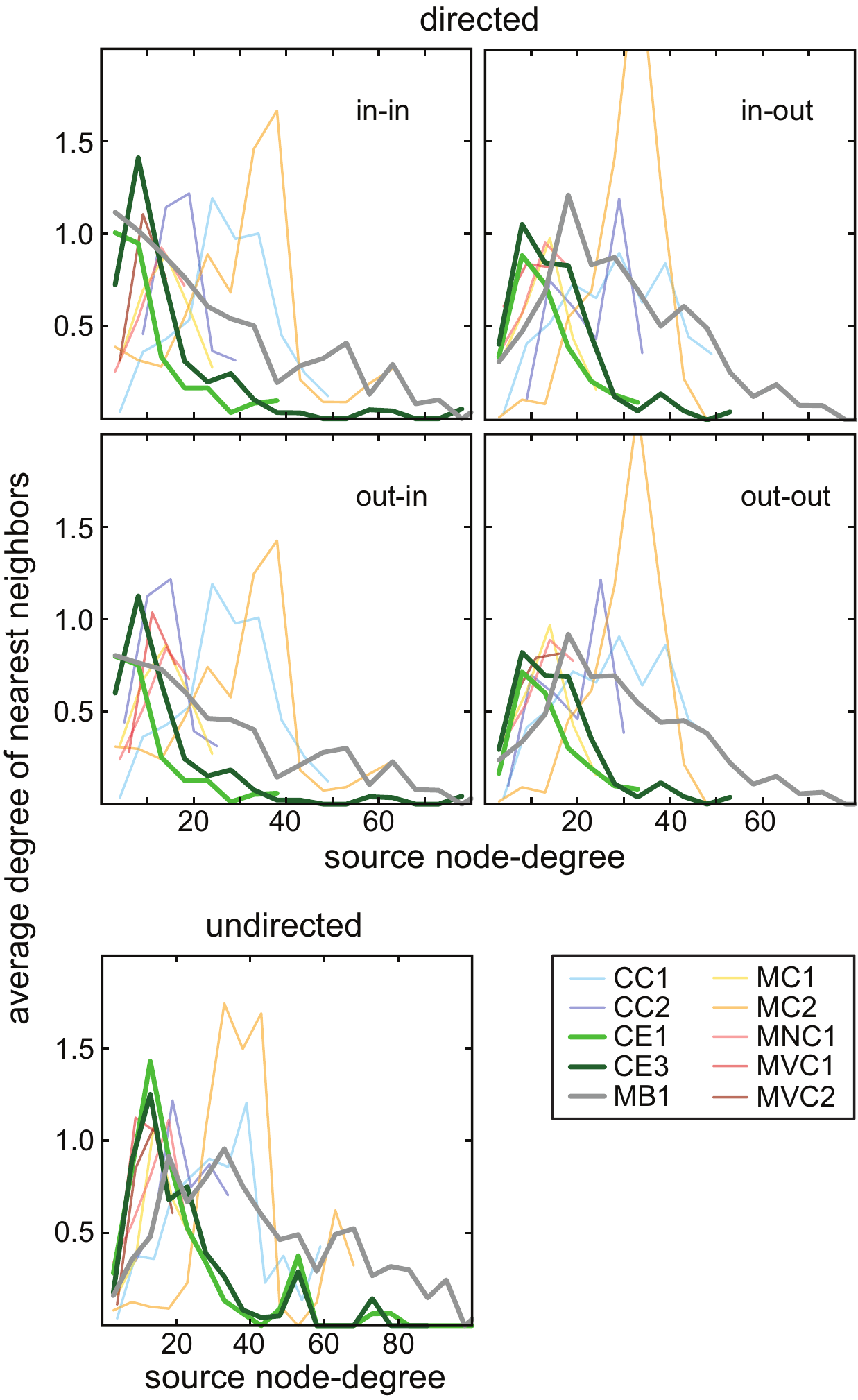}
\caption{\label{Fig_AverageDegreeOfNearestNeighbors}
Average degree of nearest neighbors for biological neural graphs. The average degree is presented as function of the source node degree (directed: {\it in-$\alpha$}: Eq.~\ref{Eq_AverageDegreeOfNNInIn}, {\it out-$\alpha$}: Eq.~\ref{Eq_AverageDegreeOfNNOutOut}; undirected graphs: Eq.~\ref{Eq_AverageDegreeOfNN}).}
\end{figure}

When considering in-edges and out-edges of the same node, however, we find that this is not the case. Here, by definition, one expects $D_{ii}^{in-out} = 0$ in undirected graphs, whereas in digraphs, if assuming a random distribution of edges, $D_{ii}^{in-out}$ should take values in accordance with Eq.~\ref{Eq_EuclideanDistanceER}. Our analysis revealed that the mean and median of $D_{ii}^{in-out}$ is significantly lower than expected from a pure random assignment of edges, leading to $\Delta^{ER} D^{\alpha-\beta}$ significantly larger than zero (Fig.~\ref{Fig_EuclideanDistances_Correlations}A, right). The latter suggests that the distribution of edges in biological neural graphs, although being consistent with a minimal random model, exhibits a significant correlation between in and out-edges for individual nodes.

\begin{table*}[t]
\caption{\label{Tab_GraphAssortativity}
Node degree correlations $r$ (Eq.~\ref{Eq_NodeDegreeCorrelation}), graph assortativity coefficient and measures $\mathfrak{r}$ (Eqs.~\ref{Eq_GraphAssortativity} and \ref{Eq_GraphAssortativityMeasure}, respectively) for various biological neural graphs. Positive values of $\mathfrak{r}$ indicate assortative mixing, negative values disassortative mixing.}
\begin{tabular*}{\textwidth}{l|p{3mm}p{15mm}|p{2mm}p{20mm}p{20mm}p{20mm}p{20mm}|p{2mm}l}
\hline
& 
& 
& \multicolumn{5}{c|}{\bf directed}
& \multicolumn{2}{c}{\bf undirected} \\
& 
& $r$
& 
& $\mathfrak{r}^{in-in}$
& $\mathfrak{r}^{in-out}$
& $\mathfrak{r}^{out-in}$
& $\mathfrak{r}^{out-out}$
&
& $\mathfrak{r}$ \\
\hline
CC1  & & 0.9515 & & -7.2935$\cdot 10^{-2}$ & -6.2895$\cdot 10^{-2}$ & -5.2792$\cdot 10^{-2}$ & -3.2209$\cdot 10^{-2}$ & & -9.1750$\cdot 10^{-2}$ \\
CC2  & & 0.7516 & & -2.5389$\cdot 10^{-2}$ &  1.5424$\cdot 10^{-2}$ & -2.7614$\cdot 10^{-2}$ &  8.2941$\cdot 10^{-2}$ & & -4.4217$\cdot 10^{-2}$ \\
CE1  & & 0.4964 & & -4.9522$\cdot 10^{-2}$ & -1.1982$\cdot 10^{-1}$ & -4.8745$\cdot 10^{-2}$ &  1.0551$\cdot 10^{-2}$ & & -1.6320$\cdot 10^{-1}$ \\
CE3  & & 0.7138 & & -7.2604$\cdot 10^{-2}$ & -9.4878$\cdot 10^{-2}$ & -9.3457$\cdot 10^{-2}$ & -7.8380$\cdot 10^{-2}$ & & -9.2295$\cdot 10^{-2}$ \\
MB1  & & 0.6828 & & -1.0062$\cdot 10^{-1}$ & -8.2594$\cdot 10^{-2}$ & -1.1031$\cdot 10^{-1}$ & -6.4468$\cdot 10^{-2}$ & & -1.3813$\cdot 10^{-1}$ \\
MC1  & & 0.8539 & &  5.5006$\cdot 10^{-2}$ &  4.5165$\cdot 10^{-2}$ & -7.6005$\cdot 10^{-3}$ & -1.1037$\cdot 10^{-2}$ & &   8.9371$\cdot 10^{-2}$ \\
MC2  & & 0.5287 & & -3.8981$\cdot 10^{-3}$ &  1.5331$\cdot 10^{-2}$ & -7.0477$\cdot 10^{-2}$ & -1.4625$\cdot 10^{-2}$ & & -1.5057$\cdot 10^{-1}$ \\
MNC1 & & 0.8336 & &  3.1641$\cdot 10^{-2}$ &  4.3503$\cdot 10^{-2}$ & -2.2502$\cdot 10^{-2}$ & -2.0529$\cdot 10^{-2}$ & &   5.6297$\cdot 10^{-3}$ \\
MVC1 & & 0.6894 & & -9.4035$\cdot 10^{-2}$ & -1.4435$\cdot 10^{-2}$ & -8.3551$\cdot 10^{-2}$ &  3.2236$\cdot 10^{-2}$ & & -2.9813$\cdot 10^{-2}$ \\
MVC2 & & 0.6894 & & -9.4035$\cdot 10^{-2}$ & -1.4435$\cdot 10^{-2}$ & -8.3551$\cdot 10^{-2}$ &  3.2236$\cdot 10^{-2}$ & & -7.5663$\cdot 10^{-2}$ \\
\hline
\end{tabular*}
\end{table*}

To further explore this point, we applied a second measure of structural equivalence, specifically the Pearson correlation coefficient of the node end-degrees \citep{BoccalettiEA06}, which can be interpreted as local assortativity (see below). For digraphs, one defines
\begin{eqnarray}
R_{ij}^{in-in} & = & N_{ij}^{in-in} \sum\limits_{k=1}^{N_N} \left( a_{ki} - \langle a_i^{in} \rangle \right) \left( a_{kj} - \langle a_j^{in} \rangle \right)
\label{Eq_CorrelationCoefficientInIn} \\
R_{ij}^{in-out} & = & N_{ij}^{in-out} \sum\limits_{k=1}^{N_N} \left( a_{ki} - \langle a_i^{in} \rangle \right) \left( a_{jk} - \langle a_j^{out} \rangle \right) \nonumber \\
& & \label{Eq_CorrelationCoefficientInOut} \\
R_{ij}^{out-out} & = & N_{ij}^{out-out} \sum\limits_{k=1}^{N_N} \left( a_{ik} - \langle a_i^{out} \rangle \right) \left( a_{jk} - \langle a_j^{out} \rangle \right)  \, , \nonumber \\
& & \label{Eq_CorrelationCoefficientOutOut}
\end{eqnarray}
where $\langle a_i^{\{in,out\}} \rangle$ denotes the mean of values in \{row,column\} $i$ of the adjacency matrix and
\begin{eqnarray*}
N_{ij}^{in-in} & = & \left\{ \sum\limits_{k=1}^{N_N} \left( a_{ki} - \langle a_i^{in} \rangle \right)^2 \left( a_{kj} - \langle a_j^{in} \rangle \right)^2 \right\}^{-1/2} \\
N_{ij}^{in-out} & = & \left\{ \sum\limits_{k=1}^{N_N} \left( a_{ki} - \langle a_i^{in} \rangle \right)^2 \left( a_{jk} - \langle a_j^{out} \rangle \right)^2 \right\}^{-1/2} \\
N_{ij}^{out-out} & = & \left\{ \sum\limits_{k=1}^{N_N} \left( a_{ik} - \langle a_i^{out} \rangle \right)^2 \left( a_{jk} - \langle a_j^{out} \rangle \right)^2 \right\}^{-1/2}
\end{eqnarray*}
are normalization constants. If self-loops are excluded, then the sum in the above equations runs over $k \neq \{i, j\}$. Moreover, as for the Euclidean distance, $[R_{ij}^{in-out}]^T = R_{ji}^{in-out} = R_{ij}^{out-in}$, so that for a given node pair $(i, j)$ only three  correlation measures are independent. It can be shown that $-1 \leq R_{ij}^{\alpha-\beta} \leq 1$. 

If two nodes $i$ and $j$ of a graph are structurally perfectly equivalent in their in-edge and out-edge distributions, i.e. share the same connection pattern with the rest of the graph, then the corresponding $R_{ij}^{\alpha-\beta} = 1$. The definitions above hold again for undirected graphs, in which case one has in addition $R_{ij}^{in-in} = R_{ij}^{out-out} = R_{ij}^{in-out}$, leaving only one independent correlation measure.

As for the Euclidean distance of node adjacencies, the correlation coefficients yield $N_N \times N_N$ matrices, and we considered the same subsets and ensemble statistics as for $D_{ij}^{\alpha-\beta}$. A representative result of the performed analysis is shown in Fig.~\ref{Fig_EuclideanDistances_Correlations}B (left) for the CC1 graph. Our analysis shows that the correlations between in-in edges $R_{ij, i \neq j}^{in-in}$, in-out edges $R_{ij}^{in-out}$ and out-out edges $R_{ij, i \neq j}^{out-out}$ (directed graphs) as well as in-in $R_{ij, i \neq j}^{in-in}$ and in-out edges $R_{ij}^{in-out}$ for undirected graphs are almost identical and close to zero, a finding which supports again a mostly uncorrelated distribution of edges across different nodes expected for random graphs (Fig.~\ref{Fig_EuclideanDistances_Correlations}B, left, red dotted). This observation is shared among all investigated biological neural graphs (Fig.~\ref{Fig_EuclideanDistances_Correlations}B, middle). However, the in-out edge distributions for the same nodes $R_{ii}^{in-out}$ show a significant positive correlation, reflecting the behavior found for Euclidean distances. This suggests that the distribution of in- and out-edges for a given node in biological digraphs is not independent (Fig.~\ref{Fig_EuclideanDistances_Correlations}C, right; for undirected graphs, $R_{ii}^{in-out} = 0$ trivially).   

Finally, the correlation between the distribution of node in- and out-edges in directed graphs was directly assessed by considering the node degree correlations. To that end, we defined the correlation coefficient of node end-degrees \citep{NewmanEA02}
\begin{equation}
\label{Eq_NodeDegreeCorrelation}
r = \frac{1}{\sigma_a^{in} \sigma_a^{out}} \left\{ \sum\limits_{n,k=1}^{\max(\Delta^{in},\Delta^{out})} ( n k p_{nk} ) - \langle a_i^{in} \rangle \langle a_i^{out} \rangle \right\} \, ,
\end{equation}
where $p_{nk}$ denotes the node degree correlations, i.e. the probability that a node with in-degree $n$ has out degree $k$, $\Delta^{\{in,out\}}$ the maximal node in/out-degree, $\sigma_a^{\{in,out\}}$ the standard deviation of node in/out-degrees
$$
\sigma_a^{\{in,out\}} = \left\{ \frac{1}{N_N} \sum\limits_{i=1}^{N_N} \left( a_i^{\{in,out\}} - \langle a_i^{\{in,out\}} \rangle \right)^2 \right\}^{1/2}
$$
and $\langle a_i^{\{in,out\}} \rangle$ the average node in/out-degree. Note that $\langle a_i^{in} \rangle = \langle a_i^{out} \rangle$. It can be shown that $-1 \leq r \leq 1$.

For undirected graphs, $r = 1$ due to the symmetry of the adjacency matrix. For the directed versions of the considered biological neural graphs, the node degree correlations are listed in Table~\ref{Tab_GraphAssortativity}. As shown, $r \gg 0$ for all considered graphs, suggesting a strong correlation between the in-degree and out-degree, with nodes of in-degree $n$ tending to have out-degree $n$. 


\section{Nearest neighbor degrees}

To further explore structural characteristics of biological neural graphs, we finally investigated the correlation between nearest neighbor degrees. For that, various measures were proposed in the literature (e.g. see \citealp{BoccalettiEA06, Newman10}). First, we calculated the average degrees of nearest neighbors, defined for digraphs as
\begin{eqnarray}
\langle a_k^{nn\ in \leftarrow \alpha} \rangle & = & \sum\limits_{n^{in}=1}^{\Delta^{in}} n^{in} p_{n^{in}|k^{\alpha}}^{in \leftarrow \alpha} \label{Eq_AverageDegreeOfNNInIn} \\
\langle a_k^{nn\ out \leftarrow \alpha} \rangle & = & \sum\limits_{n^{out}=1}^{\Delta^{out}} n^{out} p_{n^{out}|k^{\alpha}}^{out \leftarrow \alpha} \label{Eq_AverageDegreeOfNNOutOut} \, ,
\end{eqnarray}
where $\alpha \in \{in,out\}$, $p_{n^{in}|k^{\alpha}}^{in \leftarrow \alpha}$ denotes the conditional probability that an edge from a node with $\alpha$-degree $k^{\alpha}$ points to a node with in-degree $n^{in}$, and $p_{n^{out}|k^{\alpha}}^{out \leftarrow \alpha}$ the conditional probability that an edge from a node with $\alpha$-degree $k^{\alpha}$ points to a node with out-degree $n^{out}$. Similarly, for undirected graphs, one has \citep{BoccalettiEA06}
\begin{equation}
\langle a_k^{nn} \rangle = \sum\limits_{n=1}^{\Delta} n p_{n|k} \label{Eq_AverageDegreeOfNN} \, ,
\end{equation}
where $p_{n|k}$ is the conditional probability that an edge from a node of degree $k$ points to a node of degree $n$.

\begin{figure}[t]
\includegraphics[width=\columnwidth]{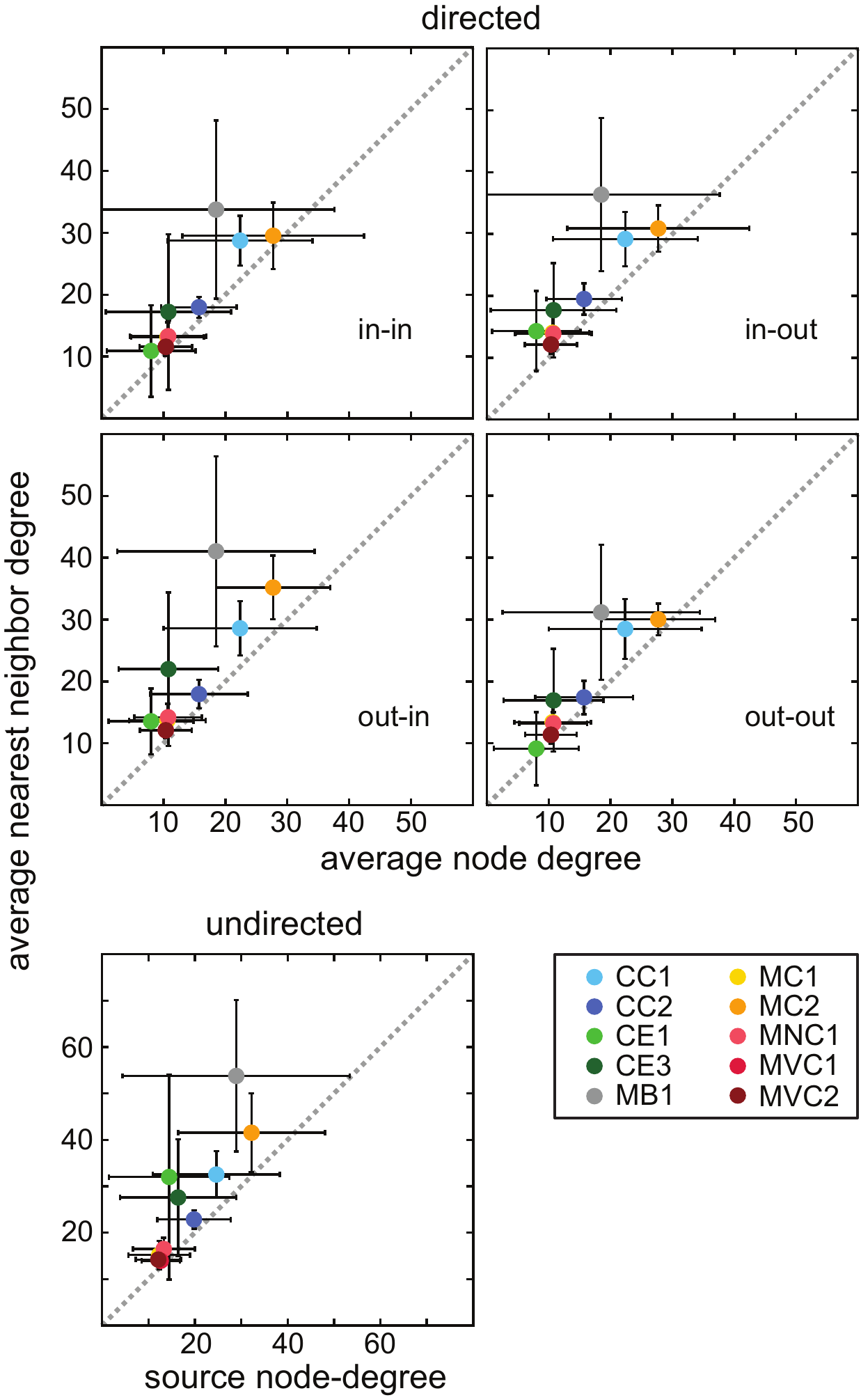}
\caption{\label{Fig_AverageNearestNeighborDegrees}
Average nearest neighbor degree for biological neural graphs. The average nearest neighbor degree is presented as function of average node degree (directed: {\it in-$\alpha$}: Eq.~\ref{Eq_AverageNearestNeighborDegreeIn}, {\it out-$\alpha$}: Eq.~\ref{Eq_AverageNearestNeighborDegreeOut}; undirected graphs: Eq.~\ref{Eq_AverageNearestNeighborDegree}.}
\end{figure}

Eqs.~\ref{Eq_AverageDegreeOfNNInIn}-\ref{Eq_AverageDegreeOfNN} yield the average degrees of nearest neighbors as function of the degree $k$ of the source node, and are visualized in Fig.~\ref{Fig_AverageDegreeOfNearestNeighbors} for the investigated neural graphs. Due to the small graph size, only graphs with the largest $N_N$ show a clear decrease of the average degree of nearest neighbors as function of the degree of the source node (Fig.~\ref{Fig_AverageDegreeOfNearestNeighbors}; {\it CE1}, {\it CE3} and {\it MB1}). This suggests that at least these graphs display a disassortative behavior, in accordance with the recent finding that biological graphs tend to be disassortative \citep{JohnsonEA10}. 

In disassortative graphs, nodes with low degree tend to be connected with nodes of high degree (and vice versa). To quantify the assortativity, we considered two additional measures. The average nearest neighbor degrees, defined for digraphs as
\begin{eqnarray}
\langle a_i^{nn\ in-\alpha} \rangle & = & \frac{1}{a_i^{in}} \sum\limits_{j=1}^{N_N} a_{ji} a_j^{\alpha} 
\label{Eq_AverageNearestNeighborDegreeIn} \\
\langle a_i^{nn\ out-\alpha} \rangle & = & \frac{1}{a_i^{out}} \sum\limits_{j=1}^{N_N} a_{ij} a_j^{\alpha} 
\label{Eq_AverageNearestNeighborDegreeOut}
\end{eqnarray}
($\alpha \in \{ in, out \}$) and for undirected graphs as
\begin{equation}
\langle a_i^{nn} \rangle = \frac{1}{a_i} \sum\limits_{j=1}^{N_N} a_{ij} a_j \, ,
\label{Eq_AverageNearestNeighborDegree}
\end{equation}
allow for a quantification of the relation between the average degree of source and target nodes. Eqs.~\ref{Eq_AverageNearestNeighborDegreeIn}-\ref{Eq_AverageNearestNeighborDegree} yield vectors of length $N_N$, and we estimated their mean value, standard deviation along with their minimum, maximum, first and third quartile as well as median. When comparing the average node in/out-degree (digraphs) and node degree (undirected graphs), one observes that in all considered graphs the average nearest neighbor degree is larger than the average node degree (Fig.~\ref{Fig_AverageNearestNeighborDegrees}). This suggests, in accordance with the finding above, that in biological neural graphs nodes tend to be connected with nodes of slightly higher average node degree. 

Finally, a direct quantification of assortative mixing is provided by the graph assortativity coefficient \citep{Newman02, Newman03}, which is equivalent to the Pearson correlation coefficient of the degree between pairs of linked nodes, and is defined by
\begin{equation}
\label{Eq_GraphAssortativity}
\mathfrak{r} = N \left\{ \sum\limits_{i=1}^{N_E} j_i k_i - \frac{1}{N_E} \sum\limits_{i=1}^{N_E} j_i \sum\limits_{i'=1}^{N_E} k_{i'} \right\} 
\end{equation}
with
\begin{eqnarray*}
N & = & \left\{ \sum\limits_{i=1}^{N_E} j_i^2 - \frac{1}{N_E} \left( \sum\limits_{i=1}^{N_E} j_i \right)^2 \right\}^{-1/2} \\
  &   & \times \left\{ \sum\limits_{i=1}^{N_E} k_i^2 - \frac{1}{N_E} \left( \sum\limits_{i=1}^{N_E} k_i \right)^2 \right\}^{-1/2} \, ,
\end{eqnarray*}
where $j_i$ and $k_i$ denote the excess degree (one less than the node degree) of the nodes that edge $i$ leads into and out of, respectively, and $N_E$ the number of graph edges. Eq.~\ref{Eq_GraphAssortativity} holds for directed and undirected graphs. In the latter case, each edge is replaced by two directed edges leading in opposite directions. Recently, another set of graph assortativity measures for digraphs was proposed, which can be viewed as a generalization of the above assortativity coefficient. It is given by \citet{FosterEA10}
\begin{equation}
\label{Eq_GraphAssortativityMeasure}
\mathfrak{r}^{\alpha-\beta} = \frac{1}{N_E \sigma^{\alpha} \sigma^{\beta}} \sum\limits_{i=1}^{N_E} \left[ \Big( j_i^{\alpha} - \overline{j^{\alpha}} \Big) \Big( k_i^{\beta} - \overline{k^{\beta}} \Big) \right] ,
\end{equation}
where
\begin{eqnarray*}
\overline{j^{\alpha}} & = & \frac{1}{N_E} \sum\limits_{i=1}^{N_E} j_i^{\alpha} \\
\overline{k^{\beta}} & = & \frac{1}{N_E} \sum\limits_{i=1}^{N_E} k_i^{\beta} \\
\sigma^{\alpha} & = & \left\{ \frac{1}{N_E} \sum\limits_{i=1}^{N_E} \left( j_i^{\alpha} - \overline{j^{\alpha}} \right)^2 \right\}^{1/2} \\
\sigma^{\beta} & = &  \left\{ \frac{1}{N_E} \sum\limits_{i=1}^{N_E} \left( k_i^{\beta} - \overline{k^{\beta}} \right)^2 \right\}^{1/2} .
\end{eqnarray*}
Here, $\alpha, \beta \in \{ in, out \}$, and $j_i^{\alpha}$, $k_i^{\beta}$ denote the $\alpha$- and $\beta$-degree of source and target node of a given edge $i$. One can show that $\mathfrak{r}$ in Eq.~\ref{Eq_GraphAssortativity} is equivalent to $\mathfrak{r}^{out-in}$ in Eq.~\ref{Eq_GraphAssortativityMeasure}. 

\begin{figure}[th!]
\includegraphics[width=\columnwidth]{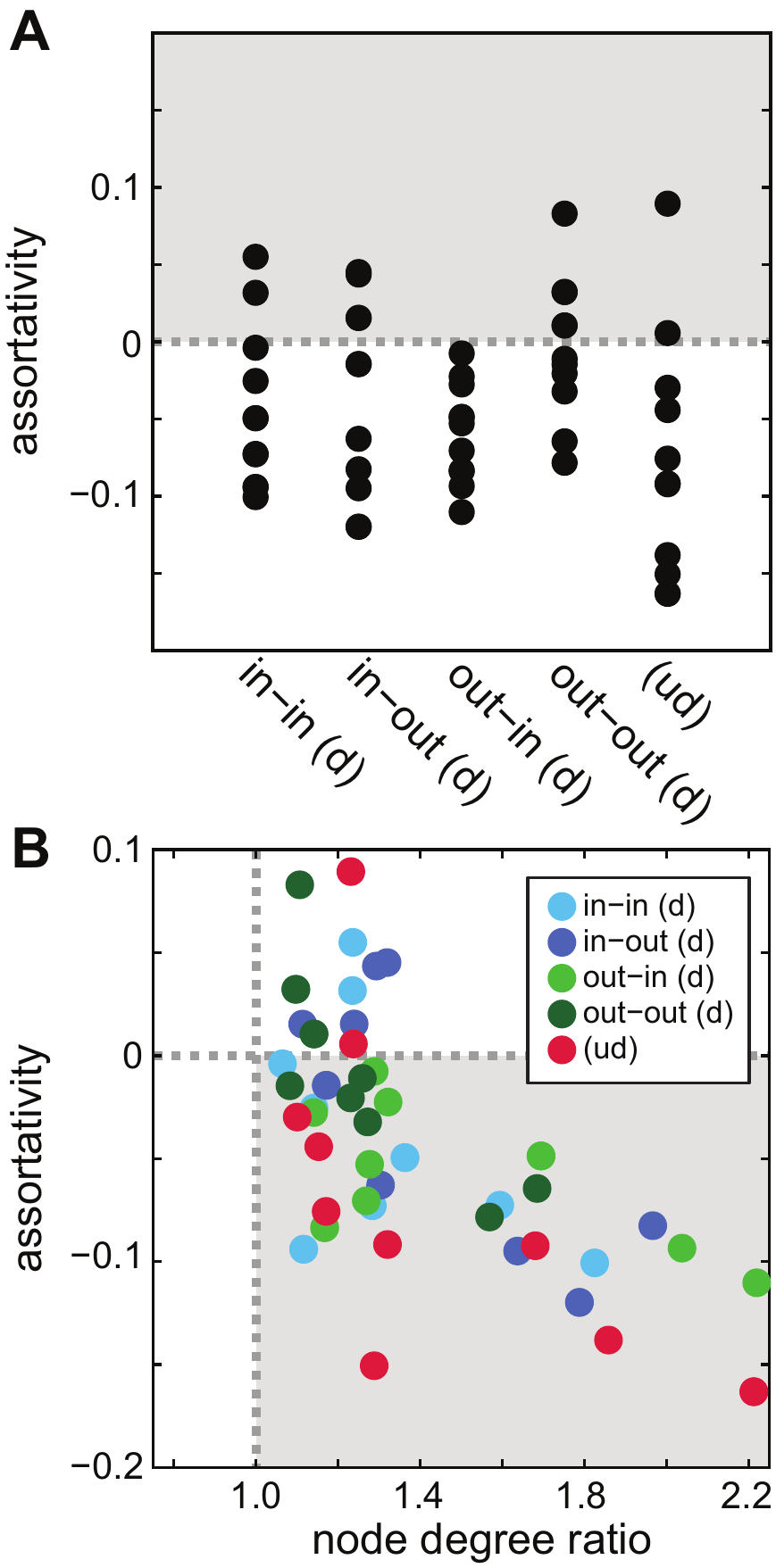}
\caption{\label{Fig_GraphAssortativity}
Graph assortativity coefficients and measures for various biological neural graphs. (a): Distribution of graph assortativity values ({\it in-in(d)}, {\it in-out(d)}, {\it out-in(d)}, {\it out-out(d)}: assortativity measures for digraphs Eq.~\ref{Eq_GraphAssortativityMeasure}; {\it (ud)}: assortativity coefficient Eq.~\ref{Eq_GraphAssortativity} for undirected graphs). The grey area indicates assortative mixing. (b): Graph assortativity as function of the node degree ratio, defined as $\langle a_i^{nn\ \alpha-\beta}\rangle / \langle a_i^{\alpha} \rangle$ ($\alpha, \beta \in \{ in, out \}$) for digraphs, and $\langle a_i^{nn}\rangle / \langle a_i \rangle$ for undirected graphs. The grey area indicates that negative assortativity is associated with nodes being connected to nodes with larger average node degree.}
\end{figure}

It can be shown that the assortativity coefficient and measures take values between -1 and 1, i.e. $-1 \leq \mathfrak{r}, \mathfrak{r}^{\alpha-\beta} \leq 1$. Graphs without assortative mixing have $\mathfrak{r}, \mathfrak{r}^{\alpha-\beta} = 0$. If $\mathfrak{r}, \mathfrak{r}^{\alpha-\beta} > 0$, the graph is said to mix assortatively, i.e. will show a bias in favor of edges between nodes with similar degree, $\mathfrak{r}, \mathfrak{r}^{\alpha-\beta} < 0$ quantifies a bias in favor of connection with dissimilar nodes (disassortative mixing). Results for the investigated biological neural graphs are displayed in Table~\ref{Tab_GraphAssortativity} and visualized in Fig.~\ref{Fig_GraphAssortativity}A.

For the majority of the investigated neural graphs, the assortativity was found to be close to but smaller than zero ($-0.1 \lesssim \mathfrak{r}, \mathfrak{r}^{\alpha-\beta} \lesssim 0$), suggesting a slight tendency for disassortative mixing, thus further confirming the findings outlined above. However, the level of disassortativity is generally weaker than reported in the literature. Finally, the relation between assortativity and node degrees was assessed by comparing the assortativity measures for each given graph with the ratio between average nearest neighbor degrees and node degree (Fig.~\ref{Fig_GraphAssortativity}B). As expected, the smaller the assortativity, the larger is the average nearest neighbor degree in relation to the average node degree of a given graph. In other words, an increase in disassortativity is associated with an increase in the dissimilarity between nodes and their nearest neighbors, a trend shared by all biological graphs considered in this study.


\section{Discussion}

In this work, we have completed a detailed comparative analysis of several networks fundamental to the application of graph theory in neuroscience, while challenging the pertinence of several established graph-theoretic concepts in the context of neural connectivity patterns. Contrary to many results reported in the neuroscientific literature, the biological graphs studied here show in many measures a consistency with randomness, as opposed to a consistency with simple models of graph construction, such the scale-free \citep{BarabasiAlbert99} graph. 

Specifically, we found that fits of the node degree distributions are in accordance with a gamma model, supporting the idea of a simple local mechanism responsible for generating neural graphs. Secondly, the Euclidean distance of node adjacencies and node degree correlations were observed to be consistent with an independent random distribution of node connections for different nodes, but with strong correlations between in-coming and out-going connections for the same node. Finally, we found a weak disassortative tendency in the considered graphs. Although the observed magnitudes are smaller than previously reported \citep{Newman02}, this disassortative mixing suggests that in neural graphs nodes tend to connect with nodes of slightly higher degree.  

This consistency with randomness is in some respects not fully surprising, as any conceivable mechanism which could give rise to such a structural make-up will be subject to fewer constraints compared to graph models conceived to fulfill a specific set of structural requirements. In particular, for scale-free graphs, various generating algorithms have been proposed, ranging from static models to evolving models more closely reflecting processes found in nature. Typically, static models construct scale-free graphs by imposing global constraints, such as the scale-free node-degree distribution itself \citep{AielloEA00, ChungLu02} or fitness \citep{GohEA01, CaldarelliEA02}. The most prominent model of evolving scale-free graphs is the classical growth and preferential attachment model (Barab\'asi-Albert model, see \citealp{AlbertEA99}), originally studied as ``Matthew effect'' \citep{Merton68} or ``cumulative advantage'' \citep{deSollaPrice65}), and its generalizations (such as the Dorogovtsev-Mendes-Samukhin, \citealp{DorogovtsevEA00} or Ravasz-Barab\'asi model \citealp{RavaszBarabasi03}). Here, the probability of linking two nodes is (linearly) proportional to the actual node degree, requiring the generating algorithm to keep track of all node degrees and, thus, non-local information about the graph at any stage of its construction. Finally, scale-free graph generation models utilizing accelerated growth \citep{DorogovtsevMendes01}, requiring the knowledge of the network size at any stage of construction, finite node memory \citep{KlemmEguiluz02}, requiring the knowledge of the activity state of each node in the graph, or duplication and divergence \citep{HohEA02, VazquezEA03}, requiring copies of arbitrarily selected graph nodes, were considered in the literature. All these models are crucially dependent on graph-wide properties in their generation algorithms, a requirement not necessary for generating graphs consistent with the observations presented in this study.

While the measures of structural similarily in this work are related to the bidirectional connectivity motif studied previously \citep{SongEA05}, our analysis uses 
a different technique for studying connectivity patterns in the network, assessing the nearest neighbor connectivity in terms of binary vectors in the adjacency matrix. Specifically, the Euclidean distance assesses the absence of the reciprocal pattern, whereas the structural correlation assesses its presence. Reciprocal connectivity patterns have been discussed previously in studies ranging from local cortical microcircuits \citep{SongEA05} to areal connectivity \citep{FellemanVanEssen91,VanEssen05}; in this work, we present the first systematic analysis to confirm the prevalence of this reciprocal connectivity in neural graphs spanning multiple spatial scales, and to exclude the generality of the other two-edge connectivity patterns in a thorough fashion.

In recent years, some studies have critiqued the validity of the random graph null hypothesis \citep{RandrupEA04}, noting specifically that the spatial nature of certain networks could confound the statistical comparison to a random graph for measures taken from real-world graphs (as in \citealp{MiloEA02}). In this work, we compare the measures for structural similarity to those of an equivalent random graph; however, we note that it is additionally possible with such an analysis to detect statistical differences between subsets of the Euclidean distance and stuctural correlation matrices (see Fig.~4, left panel). Such a comparison, free from a random graph null hypothesis, can be explored in future work.

In conclusion, we hope that, while there has been a great interest in recent years in the possibility that structural graphs share important features in common with abstract models of graph generation, in future work not only will greater care be taken in the support of such claims, but also more measurement and theory will be developed toward the discovery of new, specific graph theoretic models with explanatory power able to meet the challenges of the next-generation of large-scale experimental network reconstructions. 


\begin{acknowledgements}
The authors wish to thank OD Little for inspiring comments, and A Destexhe for continuing support. Work supported by the CNRS and the European Community (BrainScales project, FP7-269921). LM is a PhD fellow from \'Ecole des Neurosciences de Paris (ENP).
\end{acknowledgements}


\end{document}